\documentclass[a4paper,fleqn,usenatbib]{mnras}

\usepackage{newtxtext,newtxmath}
\usepackage[T1]{fontenc}
\usepackage{ae,aecompl}

\usepackage{graphicx}	% Including figure files
\usepackage{amsmath}	% Advanced maths commands
\usepackage{amssymb}	% Extra maths symbols

\usepackage{subfig}
\usepackage{xcolor}
\title[Bayesian covariance matrices]{A Bayesian method for combining theoretical and simulated covariance matrices for large-scale structure surveys}

\author[Alex Hall and Andy Taylor]{
Alex Hall$^{1}$\thanks{E-mail: ahall@roe.ac.uk} \& Andy Taylor$^{1}$
\\
$^{1}$Institute for Astronomy, University of Edinburgh, Royal Observatory, Blackford Hill, Edinburgh, EH9 3HJ, UK
}

\date{Accepted XXX. Received YYY; in original form ZZZ}

\pubyear{2018}

\begin{document}
\label{firstpage}
\pagerange{\pageref{firstpage}--\pageref{lastpage}}
\maketitle

% Abstract of the paper
\begin{abstract}
Accurate and precise covariance matrices will be important in enabling planned cosmological surveys to detect new physics. Standard methods imply either the need for many N-body simulations in order to obtain an accurate estimate, or a precise theoretical model. We combine these approaches by constructing a likelihood function conditioned on simulated and theoretical covariances, consistently propagating noise from the finite number of simulations and uncertainty in the theoretical model itself using an informative Inverse-Wishart prior. Unlike standard methods, our approach allows the required number of simulations to be less than the number of summary statistics. We recover the linear `shrinkage' covariance estimator in the context of a Bayesian data model, and test our marginal likelihood on simulated mock power spectrum estimates. We conduct a thorough investigation into the impact of prior confidence in different choices of covariance models on the quality of model fits and parameter variances. In a simplified setting we find that the number of simulations required can be reduced if one is willing to accept a mild degradation in the quality of model fits, finding that even weakly informative priors can help to reduce the simulation requirements. We identify the correlation matrix of the summary statistics as a key quantity requiring careful modelling. Our approach can be easily generalized to any covariance model or set of summary statistics, and elucidates the role of hybrid estimators in cosmological inference.
\end{abstract}

% Select between one and six entries from the list of approved keywords.
% Don't make up new ones.

%%%%%%%%%%%%%%%%%%%%%%%%%%%%%%%%%%%%%%%%%%%%%%%%%%

%%%%%%%%%%%%%%%%% BODY OF PAPER %%%%%%%%%%%%%%%%%%

\section{Introduction}
\label{sec:intro}

Planned cosmological surveys such as Euclid\footnote{\url{http://sci.esa.int/euclid/}}, LSST\footnote{\url{https://www.lsst.org/}}, and the SKA\footnote{\url{https://www.skatelescope.org/}} aim to constrain the properties of dark energy with unprecedented precision. These telescopes will observe large fractions of the sky out to high redshift, measuring many independent modes of the dark matter density field via weak gravitational lensing, galaxy clustering, and 21 cm intensity mapping.

In order for percent-level constraints on the dark energy equation-of-state to be realised, an accurate and precise determination of the posterior of cosmological parameters given the data is required. This will be obtained from a likelihood function for the observations given a cosmological model, and will depend on a model for the data as well as a covariance matrix describing the errors and correlations. Since the volumes of planned surveys is such that they can become systematics-limited, it is essential that the statistical analysis of the data is done correctly. This means an accurate characterisation of the likelihood function and its covariance matrix, in particular accounting for all sources of noise and for the leading sources of non-Gaussianity\footnote{While likelihood-free approaches also exist~\citep{10.1007/978-1-4614-3520-4_1, 2013ApJ...764..116W, 2018arXiv180507152L}, conventional likelihood methods are expected to constitute the primary analysis of planned surveys.}.

The primary constraints on new physics will come from clustering statistics such as the correlation function or power spectrum, measuring the clustering strength of tracers of the dark matter and its dependence on scale and redshift. However, modelling the mean and covariance of these statistics on small scales is challenging due to the complexities of structure formation and baryonic feedback. This is particularly true for weak lensing, since a given angular scale receives contributions from lensing on a wide range of spatial scales. While the mean and its cosmological dependence can potentially be modelled through a combination of large suites of hydrodynamic N-body simulations (e.g. \citealt{2017MNRAS.465.2936M}), emulators~\citep{2014ApJ...780..111H}, and phenomenological models (e.g. \citealt{2000ApJ...535L...9C}), modelling the covariance is more challenging due to the larger number of simulations required for validation, although this is partly mitigated by the less stringent requirements on accuracy in order to achieve unbiased parameter constraints.

Since the dimensionality of the covariance matrix is expected to be large for planned surveys (of order $10^4$, \citealt{2013MNRAS.432.1928T}), exploration of new methods which can bring down the required number of simulations is timely. The number of simulations required in standard approaches must be at least as large as the dimensionality for the covariance to be non-singular, and it is the inverse of the covariance which appears in the likelihood. Moreover, incorrect treatment of the statistics of this inverse can lead to biased and sub-optimal parameter constraints~\citep{2007A&A...464..399H, 2013MNRAS.432.1928T, 2013PhRvD..88f3537D, 2016MNRAS.456L.132S}. This suggests the number of simulations required can be very large, and new methods must be sought to bring this number down to a practical value.

In this work we focus on the problem of how to estimate the covariance matrix of cosmological summary statistics with accuracy and precision, and how to consistently propagate the statistics of this estimate through to a likelihood function and posterior. We follow the approach of~\citet{2016MNRAS.456L.132S} and modify the likelihood by marginalizing over the unknown covariance matrix, conditioned on a covariance matrix estimate. Unlike~\citet{2016MNRAS.456L.132S} however we use an informative prior on the typical values this covariance can take, using theoretical models as external information. This is strongly reminiscent of `shrinkage' estimates for the covariance~\citep{LEDOIT2004365, 2008MNRAS.389..766P, 2016MNRAS.456..278S, 2017MNRAS.466L..83J}, in which a hybrid covariance matrix estimate is formed by using a combination of a noisy but unbiased covariance estimate and a noise-free but imprecise model prediction. The standard Ledoit-Wolf estimator has the form of a weighted linear sum, with the optimal weight derived from simulations. However, this approach is problematic since the weight is itself stochastic, and this stochasticity is not correctly propagated if we just insert the inverted shrinkage estimate into a Gaussian or Student-$t$ likelihood. Hybrid covariance matrix estimates were also considered in~\citet{2013MNRAS.430.2200K}, where some confusion arose as to whether a correction was needed to de-bias the inverse of this estimate.

Here we take a Bayesian approach to combining theoretical and simulated covariance matrices, and recover the linear hybrid covariance matrix estimator as a by-product. The weight in this estimator is interpreted not as a free parameter to be optimised from simulations, but as a measure of the prior confidence one has that the unknown covariance is close to a given model. We conduct a thorough investigation of how this confidence and different choices of model affect the inference process, through their impact on the quality of model fits, the parameter posterior, and the variance of the data covariance matrix itself. Although we consider mostly toy models, we centre our analysis on the power spectrum of dark matter particles.

In Section~\ref{sec:theory} we introduce our data model and derive the marginal likelihood and its hybrid covariance matrix. We also introduce the model-fit, parameter-variance, and covariance-variance diagnostics we use to assess our choices of prior. This section contains the main result of this work, Equation~\eqref{eq:marglike}. In Section~\ref{sec:sims} we describe the simulations and models we use to test the likelihood, and present the results in Section~\ref{sec:results}, concluding in Section~\ref{sec:conc}.
  
\section{Combining theory and simulations in the likelihood}
\label{sec:theory}

In this section we will construct a probabilistic model for the typical summary statistics measured by surveys of large-scale structure, however much of the formalism is quite general. We adopt a Bayesian approach to making inferences from the data; for a review of Bayesian data analysis, see e.g.~\citet{Gelman}.

\subsection{Data model}
\label{subsec:datamodel}

We will consider a mock large-scale structure survey which measures a set of summary statistics such as the redshift-space galaxy clustering multipoles or the bispectrum of a weak lensing shear map. The full data vector $\mathbfit{y}$ could consist of auto-spectra and cross-spectra over a range of scales and redshift bins. We will also assume that the survey has access to a set of $n$ simulations for estimating the covariance matrix of the summary statistics, the estimate given by
\begin{equation}
  \hat{\mathbfss{C}} = \frac{1}{n-1}\sum_{i=1}^n(\mathbfit{X}_i - \bar{\mathbfit{X}})(\mathbfit{X}_i - \bar{\mathbfit{X}})^\intercal,
  \label{eq:covdef}
\end{equation}
where $\mathbfit{X}_i$ are the set of statistics measured from the $i^{\mathrm{th}}$ simulation realization and $\bar{\mathbfit{X}}$ is the sample mean across all realizations\footnote{Recently, \citet{2018MNRAS.473.4150F} presented a method for accurately estimating the \emph{inverse} covariance matrix (the precision matrix) in the context of large-scale structure surveys, but since the statistics of this estimator have not been thoroughly explored we will not consider estimated precision matrices further in this work.}.

We seek a likelihood for the data vector conditioned on a model $\boldsymbol{\mu}$, a covariance matrix estimate $\hat{\mathbfss{C}}$, and a theoretical model for the covariance $\mathbfss{C}_T$. We follow the approach of \citet{2009PhRvD..79h3012H} and \citet{2016MNRAS.456L.132S} (see also~\citealt{Gelman}), and marginalize over the unknown true covariance matrix $\mathbfss{C}$. The likelihood function may then be written as
\begin{equation}
  p(\mathbfit{y} | \boldsymbol{\mu}, \hat{\mathbfss{C}}, \mathbfss{C}_T) = \int \mathrm{d}\mathbfss{C} \,  p(\mathbfit{y} | \boldsymbol{\mu},  \mathbfss{C}, \hat{\mathbfss{C}},  \mathbfss{C}_T) \, p( \mathbfss{C} | \boldsymbol{\mu},  \hat{\mathbfss{C}},  \mathbfss{C}_T),
  \label{eq:likedef}
\end{equation}
where the integral is over all positive semi-definite symmetric matrices, and we have left dependencies on higher-order cumulants (such as the bispectrum) implicit. The first term in the integrand is the unmarginalized likelihood for the data vector, which can be simplified to $p(\mathbfit{y} | \boldsymbol{\mu},  \mathbfss{C})$ since the data is independent from $\hat{\mathbfss{C}}$ and $\mathbfss{C}_T$. Using Bayes' theorem we can write the second term as $ p( \mathbfss{C} | \boldsymbol{\mu},  \hat{\mathbfss{C}},  \mathbfss{C}_T) \propto  p( \hat{\mathbfss{C}} | \boldsymbol{\mu},  \mathbfss{C}) \, p(\mathbfss{C} | \boldsymbol{\mu} ,  \mathbfss{C}_T)$, which follows since the simulations are independent from the theoretical model covariance.

To make progress, we now assume that both the measured and simulated summary statistics ($\mathbfit{y}$ and $\mathbfit{X}_i$) are Gaussian-distributed. For the dark-matter power spectrum this is an excellent approximation on scales $k \lesssim 0.2 \, h \, \mathrm{Mpc}^{-1}$ down to $z \approx 0$~\citep{2000ApJ...544..597S, 2009ApJ...700..479T, 2015MNRAS.446.1756B} but begins to break down on smaller scales due to the non-linearity of the density field becoming sufficiently strong that the central limit theorem fails to be effective at driving the power spectrum to Gaussianity. On the smallest scales where noise (e.g. shape or shot noise) dominates over signal, the summary statistics become Gaussian again. The central limit theorem also breaks down on scales approaching the survey size due to the lack of sufficiently many independent modes, and in the case of two-point statistics the distribution becomes Gamma-distributed if the underlying fields are Gaussian (as they often are if the survey is sufficiently large, see e.g. \citealt{2014A&A...571A..15P}). More generally, non-Gaussianity due to the quadratic nature of power spectra is important for parameter inference whenever the number of independent modes per $k$-bin is not much larger than the total number of bins~\citep{2008PhRvD..77j3013H}. Similar statements can be made about two-point statistics in weak lensing~\citep{2009A&A...504..689H, 2018MNRAS.473.2355S}. Other kinds of summary statistic measured by cosmological surveys, such as the group multiplicity function,  can also possess non-Gaussian distributions~\citep{2018arXiv180306348H}. Nevertheless, to make analytic progress, we will assume Gaussian distributions for $\mathbfit{y}$ and $\mathbfit{X}_i$, noting only that any distribution may be inserted into Equation~\eqref{eq:likedef} if one is willing to perform the marginalization numerically.

With the summary statistics assumed to be Gaussian distributed, the distribution of the estimated covariance matrix given the true mean and covariance $ p( \hat{\mathbfss{C}} | \boldsymbol{\mu},  \mathbfss{C})$ is Wishart (see, e.g. \citealt{Gupta}) with $n-1$ degrees of freedom and scale matrix $\mathbfss{C}/(n-1)$, i.e. $\hat{\mathbfss{C}} \sim W_p[\mathbfss{C}/(n-1), n-1]$ where $p$ is the length of the data vector. This distribution is independent of $\boldsymbol{\mu}$ and has a probability density given by
\begin{equation}
  p(\hat{\mathbfss{C}} | \mathbfss{C}) = \frac{\lvert \hat{\mathbfss{C}} \rvert^{(n-p-2)/2} \, \mathrm{e}^{-\frac{1}{2}\mathrm{Tr}\left[(n-1)\hat{\mathbfss{C}}\mathbfss{C}^{-1}\right]}}{\lvert 2\mathbfss{C}/(n-1) \rvert^{(n-1)/2} \, \Gamma_p \left(\frac{n-1}{2} \right)},
\end{equation}
where $\Gamma_p$ is the multivariate Gamma function, and we require $n \geq p+1$. With $p(\mathbfit{y} | \boldsymbol{\mu},  \mathbfss{C})$ taken to be a $p$-dimensional multivariate Gaussian, the only distribution left to specify is the prior on the covariance given the model, $p(\mathbfss{C} | \boldsymbol{\mu} ,  \mathbfss{C}_T)$.

\subsection{Choosing the prior}
\label{subsec:priors}

There are several considerations in choosing the prior distribution on the unknown covariance matrix given the model, $p(\mathbfss{C} | \boldsymbol{\mu} ,  \mathbfss{C}_T)$. Firstly, it is highly desirable to have a prior which allows the marginalization in Equation~\eqref{eq:likedef} to be performed analytically. The alternative requires a high-dimensional integral to be performed, requiring the $\sim p^2$ free parameters of the covariance matrix to be included in the final MCMC chains, with a large associated increase in the computational complexity of the inference procedure. Secondly, our model for the covariance matrix informs the plausible values it may take, and we wish to fold in this information into our marginal likelihood. The alternative to this would be to take an uninformative (e.g. Jeffreys') prior on $\mathbfss{C}$, which is the approach taken by~\citet{2016MNRAS.456L.132S}. The Jeffreys' prior in this case is $p(\mathbfss{C} | \boldsymbol{\mu} ,  \mathbfss{C}_T) \propto \lvert \mathbfss{C} \rvert^{-(p+1)/2}$, and the resulting likelihood is a multivariate Student-$t$ distribution. However, this discards the information we have from our model for the covariance matrix, and does not permit the number of simulations to be less than $p + 1$.

A distribution which satisfies both of the above requirements is the Inverse Wishart (IW) distribution (e.g.~\citealt{Gelman}), with density
\begin{equation}
  p(\mathbfss{C} | \mathbf{\Psi}, m) = \frac{\lvert \mathbf{\Psi}/2 \rvert^{m/2}}{\Gamma_p\left(\frac{m}{2}\right)} \lvert \mathbfss{C} \rvert^{-(m+p+1)/2} \, \mathrm{e}^{-\frac{1}{2}\mathrm{Tr}(\mathbf{\Psi} \, \mathbfss{C}^{-1})},
\end{equation}
where $\mathbf{\Psi}$ is a positive-definite scale matrix and $m$ is a degree-of-freedom parameter, and we require $m > p-1$. The mean of this distribution is $\mathbf{\Psi}/(m-p-1)$ (for $m > p+1$) and its mode is $\mathbf{\Psi}/(m+p+1)$ (see, e.g. \citealt{Gupta}). We will choose the scale matrix such that the mean of the prior is equal to the theoretical model $\mathbfss{C}_T$, i.e. $\mathbf{\Psi} = (m-p-1)\mathbfss{C}_T$, and we now require $m > p+1$. This is a somewhat arbitrary choice, as we could just have well have fixed $\mathbf{\Psi}$ such that the mode was equal to $\mathbfss{C}_T$. Both are equivalent when $m \gg p$, and the reader should bear in mind this choice when $m \gtrsim p-1$.

With this choice of mean, the prior standard deviation of the $(i,j)$ element of $\mathbfss{C}$ in units of the mean is
\begin{equation}
  f_P^{ij} \equiv \frac{\sqrt{\mathrm{var}(\mathbfss{C}_{ij})}}{\langle \mathbfss{C}_{ij} \rangle} = \sqrt{\frac{(m-p+1) + (m-p-1)/\rho^2_{T,ij}}{(m-p)(m-p-3)}},
  \label{eq:fpij}
\end{equation}
where $\rho_{T,ij} = \mathbfss{C}_{T,ij}/\sqrt{\mathbfss{C}_{T,ii}\mathbfss{C}_{T,jj}}$ and we require $m > p+3$. Note that this quantity can potentially diverge when off-diagonal elements in $\mathbfss{C}_T$ are small. The standard deviation on the diagonal elements in units of the mean is then
\begin{equation}
  f_P^{ii} \equiv \frac{\sqrt{\mathrm{var}(\mathbfss{C}_{ii})}}{\langle \mathbfss{C}_{ii} \rangle} = \sqrt{\frac{2}{(m-p-3)}}.
  \label{eq:fpdef}
\end{equation}
We thus see that the degree-of-freedom parameter $m$ (the `hyperparameter') controls the `width' of the distribution. Comparison of Equation~\eqref{eq:fpdef} with the equivalent expression for the standard covariance estimate identifies the combination $m-p-2$ as an `effective number of simulations' quantifying the information brought by the prior. When $m \gg p$ the distribution tightens around the model $\mathbfss{C}_T$, becoming a delta-function in the limit $m \rightarrow \infty$. Since it is the value of $m-p$ which appears everywhere in the moments of the IW distribution, we will often refer to $f_P^{ii}$ as the free parameter of the prior in favour of $m$. This provides a useful way of controlling the confidence one has in the model, prior to running any suite of simulations to get an estimate of $\mathbfss{C}$ or seeing any data. A high value of $f_P^{ii}$ indicates a low confidence in the model covariance, with the corresponding prior broad (but still informative), while a low value of $f_P^{ii}$ indicates a high degree of confidence in the model covariance.

The choice of $f_P^{ii}$ should be specified along with the model, and could be chosen for example by comparing the model with a set of low-precision simulations (separate to those used to form the estimate $\hat{\mathbfss{C}}$ in the likelihood). If one was to find that the model diagonal elements agreed with these simulations to some precision (i.e. to some standard deviation of the simulation estimates in units of the model prediction), one could set $f_P^{ii}$ equal to that precision. If instead the model was inconsistent with the simulations by some amount of standard deviations, one could broaden the prior by choosing a higher value of $f_P^{ii}$ - as we shall see, this down-weights the influence of the prior in the marginal likelihood. Alternatively one could perform similar tests with off-diagonal elements or correlation matrices, with as broad a prior as possible chosen to bracket the uncertainties if one wished to be conservative.

For example, in~\citet{2009ApJ...700..479T}, the halo model prediction for the covariance matrix is compared to a suite of N-body simulations. At $z=0$, the halo model over-predicts the diagonal elements on scales $k \gtrsim 0.25 \, h^{-1} \, \mathrm{Mpc}$ (see their Figure 1), with the prediction being too high by roughly 30\% at $k \gtrsim 0.35 \, h^{-1} \, \mathrm{Mpc}$, with the caveat that the measurements in~\citet{2009ApJ...700..479T} have no error bars due to the finite of number of simulations available. The correlation matrix is also overestimated, by roughly 10\% on these scales (see their Figure 2). A rough way to acknowledge this imperfection in the halo model would just be to bracket these biases by setting $f_P^{ii} = 0.30$ for the halo model at $z=0$. Of course, one is free to have more confidence in the halo model if one wishes, at the cost of up-weighting these imperfections in the data covariance.

Some approaches to modelling $\hat{\mathbfss{C}}$ such as those based on effective field theory or response functions~\citep{2017JCAP...11..051B} can also produce estimates for the error of their predictions (for example by considering the impact of neglected higher-order terms), which could be used in the IW prior. Another approach would be to treat $f_P^{ii}$ as a free parameter to be marginalized over as part of a hierarchical model. In this case one would specify a prior on $f_P^{ii}$ (or $m$) and include the hyperparameter(s) in the MCMC chain along with the other cosmological and nuisance parameters, which adds very little extra computational complexity to the inference process.

One of the main aims of this work is to investigate how different choices of $\mathbfss{C}_T$ and $f_P^{ii}$ influence the final model fits and parameter constraints. While there is a certain degree of arbitrariness in how one chooses $f_P^{ii}$, we believe this improves upon previous approaches which have implicitly assumed either $f_P^{ii} = 0$ (theory-only covariance) or $f_P^{ii} = \infty$ (simulation-only covariance). We view the freedom to choose $f_P^{ii}$ in an educated way as a benefit of our approach.

Before deriving the marginal likelihood obtained by integrating out $\mathbfss{C}$, we should point out some disadvantages in the choice of an IW prior. Firstly, our choice is strongly motivated by making the resulting marginalization tractable, and not by more careful considerations of how prior information on the covariance matrix should be expressed in a prior density. As noted above, any choice for the prior can be used if one is willing to perform the marginalization numerically. Secondly, there is very little freedom in the IW distribution, with only a single free parameter once the mean has been fixed. In particular this leads to dependencies between correlations and variances~\citep{Alvarez}. Additionally, our confidence in the model is really a function of scale and redshift - on large scales or at high redshifts we might expect that perturbation theory well-describes the dominant contributions, and so our confidence here should be higher, decreasing as we push to smaller scales where the model might break down. On very small scales the covariance might be dominated by shot or shape noise, and again our confidence in the model will become high. The IW prior does not have enough freedom to capture these variations, although the one-parameter model is still an improvement over alternative choices which do not incorporate any information at all from the model. In Appendix~\ref{app:IWmix} we explore the idea of mixing IW distributions to account for this deficiency.

\subsection{Marginal likelihood}
\label{subsec:marglike}

With the choice of an IW prior, we can perform the marginalization in Equation~\eqref{eq:likedef} analytically by recasting the integrand as a new IW distribution and using its normalization property. The result is
\begin{equation}
  p(\mathbfit{y} | \boldsymbol{\mu}, \hat{\mathbfss{C}}, \mathbfss{C}_T) = \frac{\Gamma \left(\frac{\nu + p}{2}\right) \lvert \mathbfss{C}_{\mathbfit{y}} \rvert^{-1/2}}{\Gamma \left(\frac{\nu}{2}\right) [\pi(\nu-2)]^{p/2}} \left[ 1 + \frac{(\mathbfit{y} - \boldsymbol{\mu})^\intercal \mathbfss{C}_{\mathbfit{y}}^{-1}(\mathbfit{y} - \boldsymbol{\mu})}{\nu-2}\right]^{-\frac{(\nu+p)}{2}},
  \label{eq:marglike}
\end{equation}
where we have defined the quantities
\begin{align}
  &\nu \equiv n + m - p, \nonumber\\
  &\mathbfss{C}_{\mathbfit{y}} \equiv \frac{(n-1)\hat{\mathbfss{C}} + (m-p-1)\mathbfss{C}_T}{n+m-p-2}.
  \label{eq:Cy}
\end{align}
Equation~\eqref{eq:marglike} is the main result of this work, and represents a tractable analytic likelihood function incorporating both simulation-based and theory-based covariance matrix estimates. The reader is reminded at this stage that $m$ denotes the degree-of-freedom parameter of the IW prior, $n$ is the number of simulations entering the covariance matrix estimate $\hat{\mathbfss{C}}$, and $p$ is the dimensionality of the data vector. Note that since we require $m > p+1$, the matrix $\mathbfss{C}_{\mathbfit{y}}$ is positive semi-definite. The marginal likelihood is thus a multivariate $t$-distribution, with degree-of-freedom parameter $\nu$, location parameter $\boldsymbol{\mu}$ and scale matrix $(\nu-2)\mathbfss{C}_{\mathbfit{y}}/\nu$. A hybrid covariance matrix similar to Equation~\eqref{eq:Cy} was also independently derived in~\citet{2009PhRvD..79h3012H} in the context of cosmic microwave background analysis\footnote{We thank Antony Lewis for pointing out this reference.}.

The mean of the marginal likelihood is $\boldsymbol{\mu}$ and the covariance matrix, which we will refer to henceforth as the \emph{data covariance}, is $\mathbfss{C}_{\mathbfit{y}}$. The data covariance may be written as $\mathbfss{C}_{\mathbfit{y}} = (1-\lambda) \hat{\mathbfss{C}}  + \lambda\mathbfss{C}_T$, with $\lambda \equiv (m-p-1)/(n+m-p-2)$. We thus see that the parameter combination $m-p-1$, directly related to the prior width through Equation~\eqref{eq:fpdef}, controls the relative contribution of the model covariance to the total data covariance. In the limit $m - p - 1 \gg n - 1 $ we have $\mathbfss{C}_{\mathbfit{y}} \rightarrow \mathbfss{C}_T$, i.e. the data covariance `shrinks' to the model, while in the opposite limit  $n-1 \gg m-p-1 $ the noise from the simulation-based estimate is low enough that  $\mathbfss{C}_{\mathbfit{y}} \rightarrow \hat{\mathbfss{C}}$. We have effectively recovered a form of the linear shrinkage covariance estimate, but now properly embedded within a Bayesian data model. This avoids the need to estimate the shrinkage coefficient $\lambda$ from the simulations (as suggested in \citealt{LEDOIT2004365, 2008MNRAS.389..766P}), and hence avoids introducing unaccounted-for noisy estimates into the likelihood.

The marginal likelihood in Equation~\eqref{eq:marglike} incorporates information on the covariance matrix from both the simulations and the theoretical model. By contrast, had we assumed an uninformative Jeffreys' prior on $\mathbfss{C}$, the data covariance matrix would simply be given by~\citep{2016MNRAS.456L.132S}\footnote{The marginal likelihood assuming a Jeffreys' prior can be found from Equation~\eqref{eq:marglike} by taking $\mathbfss{C}_T = 0$ and $m=0$, with the condition that $n \geq p + 1$.}
\begin{equation}
  \mathbfss{C}^{J}_{\mathbfit{y}} =  \frac{n-1}{n-p-2}\hat{\mathbfss{C}},
  \label{eq:CyJ}
\end{equation}
where $n \geq p+1$. Note that the Jeffreys' data covariance matrix is always larger elementwise than the estimated covariance since the marginalization over an uninformative prior has broadened the likelihood. This is in contrast to the IW data covariance in Equation~\eqref{eq:Cy}, which can be smaller or large than $\hat{\mathbfss{C}}$ depending on how much weight is assigned to the model.

Finally, an important point to note is that with an IW prior we do not require $n \geq p+1$ for $\mathbfss{C}_{\mathbfit{y}}$ to be full rank, due to the regularizing influence of the prior\footnote{This follows from the subadditivity law of matrix ranks.}. This allows the number of simulations to be smaller than the number of summary statistics, raising hopes that we might significantly reduce the computational resources required to build the likelihood with this approach.

\subsection{Sampling distribution of the data covariance matrix}
\label{subsec:Cystats}

The data covariance matrix $\mathbfss{C}_{\mathbfit{y}}$ in Equation~\eqref{eq:Cy} has a complicated sampling distribution which resembles a shifted Wishart distribution. In particular, its mean is biased with respect to the true value $\mathbfss{C}_0$ if the model is not equal to the truth. This is not a problem per se, since the sampling distribution is not the relevant quantity here; what matters is the marginal likelihood (and ultimately the posterior on the cosmological parameters) conditioned on the estimated and model covariance matrices. Nevertheless, it is clearly desirable to have a data covariance matrix which is on average close to the true value - to achieve this we either require an accurate model with a high weight within $\mathbfss{C}_{\mathbfit{y}}$, or a large number of simulations. In this subsection we will investigate the statistics of the data covariance.

The higher-order cumulants of $\mathbfss{C}_{\mathbfit{y}}$ are identical to those of a Wishart-distributed matrix with scale matrix $\mathbfss{C}_0/(n+m-p-2)$ and degree-of-freedom parameter $n-1$. In particular, the variance of the elements of $\mathbfss{C}_{\mathbfit{y}}$ is
\begin{equation}
  \mathrm{var}\left(\mathbfss{C}_{\mathbfit{y},ij}\right) = \frac{n-1}{(n+m-p-2)^2}\left(\mathbfss{C}_{0,ii}\mathbfss{C}_{0,jj} + \mathbfss{C}_{0,ij}^2 \right),
  \label{eq:varCy}
\end{equation}
which is to be compared with the equivalent for the covariance matrix estimate $\hat{\mathbfss{C}}$
\begin{equation}
  \mathrm{var}\left(\hat{\mathbfss{C}}_{ij}\right) = \frac{1}{(n-1)}\left(\mathbfss{C}_{0,ii}\mathbfss{C}_{0,jj} + \mathbfss{C}_{0,ij}^2 \right).
  \label{eq:varChat}
\end{equation}
Differentiating Equation~\eqref{eq:varCy} with respect to $n$, we find that the data covariance has a maximum when $n=m-p$, i.e. when equal weight is assigned to the simulations and theoretical model. For larger values of $n$ the simulation-based estimate converges to the truth (albeit slowly), while for smaller $n$ the zero-variance theoretical model gets more weight.

\begin{figure}
  \includegraphics[width=\columnwidth]{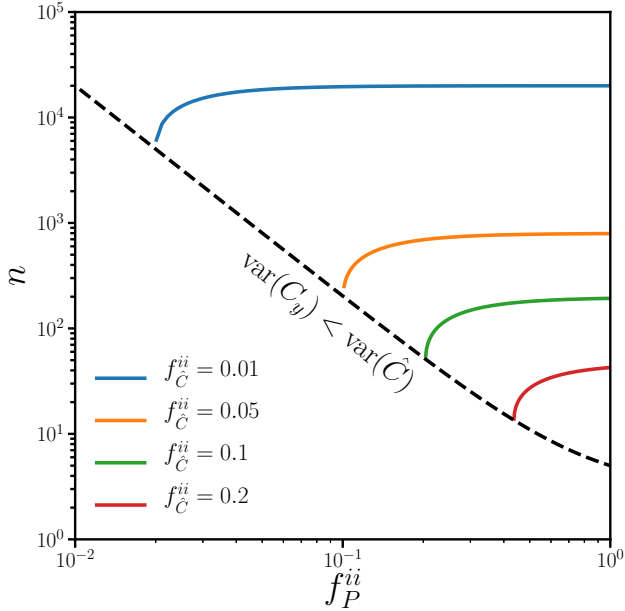}
  \caption{The number of simulations $n$ which the data covariance matrix needs in order to match the uncertainty of the standard covariance matrix estimate, as a function of the width of the theory prior $f_P^{ii}$ (defined in Equation~\eqref{eq:fpdef}). The uncertainty in the standard estimate is labelled by $f_{\hat{\mathbfss{C}}}^{ii}$ (the standard deviation of the diagonal elements in units of the mean), for $f_{\hat{\mathbfss{C}}}^{ii} = 0.01$ (blue, top curve),  $f_{\hat{\mathbfss{C}}}^{ii} = 0.05$ (orange, upper middle curve),  $f_{\hat{\mathbfss{C}}}^{ii} = 0.1$ (green, lower middle curve), and  $f_{\hat{\mathbfss{C}}}^{ii} = 0.2$ (red, lower curve). The black dashed line shows the critical value $n=m-p$. Choices of $n$ and $f_P^{ii}$ to the bottom-left of this line always achieve lower variance than the standard estimate.}
  \label{fig:preservevarJ}
\end{figure}

As well as requiring the mean of the data covariance $\mathbfss{C}_{\mathbfit{y}}$ to be close to the true value, it is also clearly desirable to keep the variance of $\mathbfss{C}_{\mathbfit{y}}$ reasonably low.  Given some specified precision on the standard covariance estimate $\hat{\mathbfss{C}}$, we can determine how many simulations need to be run in order to match that precision for different choices of the prior width $f_P^{ii}$. We do this by equating Equation~\eqref{eq:varCy} with Equation~\eqref{eq:varChat} and solving the resulting quadratic equation. In Figure~\ref{fig:preservevarJ} we plot the number of simulations required for the variance of $\mathbfss{C}_{\mathbfit{y}}$ to match that of $\hat{\mathbfss{C}}$ as a function of the prior width $f_P^{ii}$ and for different values of the precision $f_{\hat{\mathbfss{C}}}^{ii}$, defined as the standard deviation of the diagonal elements of $\hat{\mathbfss{C}}$ in units of the mean. For a given requirement on the precision of the diagonal elements (10\% say, corresponding to 200 simulations if no extra information is included, and the red curve in Figure~\ref{fig:preservevarJ}), we can reduce the number of simulations required by lowering $f_P^{ii}$, i.e. by using a more informative prior. In the limit of a very broad prior (a large value of $f_P^{ii}$), we do not gain anything and $n$ asymptotes to the value required for the standard estimate to have the given precision. As we reduce $f_P^{ii}$ we eventually reach a critical value of $n$ below which the variance on $\mathbfss{C}_{\mathbfit{y}}$ is necessarily \emph{lower} than the variance of $\hat{\mathbfss{C}}$. This regime is reached for either a small value of $n$ or small values of $f_P^{ii}$, both corresponding to high weight on the noiseless theoretical model.

In summary, Figure~\ref{fig:preservevarJ} demonstrates, with no assumptions about the form of $\mathbfss{C}_T$ or $\mathbfss{C}_0$, that we can achieve similar precision on the data covariance matrix to the standard approach but with fewer simulations. Although this is clearly desirable, there are other considerations at play when determining the number of simulations that must be run, in particular the effects of reducing $n$ on the variance of model parameters and the goodness-of-fit of the best-fitting model.

\subsection{Posterior on model parameters}
\label{subsec:post}

Equipped with the marginal likelihood of Equation~\eqref{eq:marglike}, we can use Bayes' theorem to determine the posterior on a set of model parameters $\boldsymbol{\theta}$. A straightforward first step in studying this posterior is to approximate its covariance matrix with the Fisher matrix, whose $(\alpha,\beta)$ element is
\begin{equation}
  \mathbfss{F}_{\alpha \beta} = \left \langle \frac{\partial \ln{P}}{\partial \theta_\alpha}\frac{\partial \ln{P}}{\partial \theta_\beta}\right \rangle,
\end{equation}
where the angle brackets denote an expectation over the marginal likelihood, the partial derivatives are with respect to the parameters, and $P$ is the marginal likelihood. For simplicity we will assume that only the mean $\boldsymbol{\mu}(\boldsymbol{\theta})$ depends on the parameters, in which case the expectation is straightforward (see, e.g.~\citealt{2017MNRAS.464.4658S}) and gives
\begin{equation}
  \mathbfss{F}_{\alpha \beta} = \frac{(n+m)(n+m-p)}{(n+m+2)(n+m-p-2)} \frac{\partial \boldsymbol{\mu}^\intercal}{\partial \theta_\alpha} \mathbfss{C}_{\mathbfit{y}}^{-1} \frac{\partial \boldsymbol{\mu}}{\partial \theta_\beta}.
  \label{eq:fisher}
\end{equation}
This expression agrees with the Gaussian result in the limits $n \rightarrow \infty$ and $m \rightarrow \infty$, where the data covariance tends to the true covariance and the model covariance respectively.

The Fisher matrix in Equation~\eqref{eq:fisher} has been averaged over realizations of the data (under the marginal likelihood) but still depends on the particular realization of the simulation estimate $\hat{\mathbfss{C}}$.  Further averaging over the sampling distribution of $\hat{\mathbfss{C}}$ allows one to study the loss of Fisher information on parameters resulting from a finite number of simulations (see Figure 6 of~\citealt{2017MNRAS.464.4658S}). Unfortunately the presence of $\mathbfss{C}_T$ in the data covariance matrix precludes us from analytically computing the expectation value of the Fisher matrix (or its inverse). To make progress we specialize to a single parameter, which we take as the amplitude $A$ of the mean, with the model specified as
\begin{equation}
  \boldsymbol{\mu}(A) = A \boldsymbol{\mu}_0,
  \label{eq:mumodel}
\end{equation}
where $\boldsymbol{\mu}_0$ is a fiducial model corresponding to the value $A=1$. Assuming a flat prior on $A$ it is easy to show that $A$ is distributed as Student-$t$ with posterior mean and variance given by
\begin{align}
  &\langle A \rangle = \frac{\boldsymbol{\mu}_0^\intercal \mathbfss{C}_{\mathbfit{y}}^{-1}\mathbfit{y}}{\boldsymbol{\mu}_0^\intercal \mathbfss{C}_{\mathbfit{y}}^{-1} \boldsymbol{\mu}_0}, \nonumber \\
  &\mathrm{var}(A) = \frac{(n+m-p-2)}{(n+m-3)\boldsymbol{\mu}_0^\intercal \mathbfss{C}_{\mathbfit{y}}^{-1} \boldsymbol{\mu}_0} \left[1 + \frac{\tilde{\mathbfit{y}}^\intercal \tilde{\mathbfss{C}}^{-1}_{\mathbfit{y}} \tilde{\mathbfit{y}}}{(n+m-p-2)}\right],
  \label{eq:Astats}
\end{align}
where $\tilde{\mathbfit{y}}$ and $\tilde{\mathbfss{C}}_{\mathbfit{y}}$ are the data vector and data covariance projected orthogonal to $\boldsymbol{\mu}_0$ respectively. The quantity in the numerator in the square brackets in Equation~\eqref{eq:Astats} is given by
\begin{equation}
  \tilde{\mathbfit{y}}^\intercal \tilde{\mathbfss{C}}^{-1}_{\mathbfit{y}}\tilde{\mathbfit{y}} =  \mathbfit{y}^\intercal \left( \mathbfss{C}_{\mathbfit{y}}^{-1} - \frac{ \mathbfss{C}_{\mathbfit{y}}^{-1} \boldsymbol{\mu}_0 \boldsymbol{\mu}_0^\intercal \mathbfss{C}_{\mathbfit{y}}^{-1}}{\boldsymbol{\mu}_0^\intercal \mathbfss{C}_{\mathbfit{y}}^{-1} \boldsymbol{\mu}_0} \right) \mathbfit{y}.
  \label{eq:projCy}
\end{equation}
Note that unlike the Gaussian case, the posterior variance on the amplitude depends on the data through the term in square brackets in Equation~\eqref{eq:Astats}, with the dependence vanishing when $n$ becomes large. This contribution does not appear in the elements of the inverse Fisher matrix, where non-Gaussianity in the posterior is neglected. Note also that further averaging over the data gives $\langle A \rangle = A_0$ where $A_0$ is the true value of $A$, i.e. the mean of the posterior (which is both the maximum likelihood estimate and the maximum a posteriori estimate in our case) is unbiased for any choice of $\mathbfss{C}_T$ (c.f. ~\citealt{2013PhRvD..88f3537D,2015JCAP...12..058W}). %In Appendix~\ref{app:parbias} we show that this property holds for any parameter entering only through the mean $\boldsymbol{\mu}(\boldsymbol{\theta})$.

Finally we note that averaging the variance in Equation~\eqref{eq:Astats} over the data is straightforward since it is quadratic in the data vector, but further averaging over the realizations of $\hat{\mathbfss{C}}$ is non-trivial and must be performed numerically.

\subsection{Model fitting with the marginal likelihood}
\label{subsec:PTEs}

Assuming the simple amplitude model specified by Equation~\eqref{eq:mumodel}, we can use our marginal likelihood to derive the posterior distribution of $A$. A natural best-fitting model may be found by taking the mean (or equivalently the maximum) of this posterior, given in terms of the data by the first line of Equation~\eqref{eq:Astats}. We will denote by $\hat{A}$ this best-fitting estimate of the model amplitude. Once this best-fit has been derived, a rough measure of the quality of the fit can be made by constructing a test statistic from $\hat{A}$. We then compute the probability of obtaining a value at least as big as this test statistic assuming it obeys a distribution specified by a null hypothesis - the `probability to exceed' (PTE)\footnote{This terminology is non-standard outside of the cosmic microwave background literature, but we prefer it over `$p$-value' to avoid confusion over whether we compute one-sided or two-sided $p$-values. The reader should bear in mind that with PTE we are referring to a one-sided $p$-value.}. An obvious choice for the test statistic is something proportional to the sum of the squared residuals, given by
\begin{align}
  \chi^2 &= \left( \mathbfit{y} - \hat{A}\boldsymbol{\mu}_0\right)^\intercal \mathbfss{C}_{\mathbfit{y}}^{-1} \left(\mathbfit{y} - \hat{A}\boldsymbol{\mu}_0 \right) \nonumber \\
  &= \mathbfit{y}^\intercal \left( \mathbfss{C}_{\mathbfit{y}}^{-1} - \frac{ \mathbfss{C}_{\mathbfit{y}}^{-1} \boldsymbol{\mu}_0 \boldsymbol{\mu}_0^\intercal \mathbfss{C}_{\mathbfit{y}}^{-1}}{\boldsymbol{\mu}_0^\intercal \mathbfss{C}_{\mathbfit{y}}^{-1} \boldsymbol{\mu}_0} \right) \mathbfit{y} \nonumber \\
  &= \tilde{\mathbfit{y}}^\intercal \tilde{\mathbfss{C}}^{-1}_{\mathbfit{y}}\tilde{\mathbfit{y}},
  \label{eq:chi2}
\end{align}
where in the second line we have used the definition of $\hat{A}$ and used the definition of the projected data covariance $\tilde{\mathbfss{C}}_{\mathbfit{y}}$ and data vector $\tilde{\mathbfit{y}}$ in the final line, see Equation~\eqref{eq:projCy}. These projected quantities are defined in the $p-1$ dimensional hypersurface orthogonal to $\boldsymbol{\mu}_0$. In particular, the sampling distribution of $\tilde{\mathbfit{y}}$ is a $p-1$ dimensional multivariate Gaussian with zero mean and covariance $\tilde{\mathbfss{C}}_0$.

We still need to specify a null hypothesis from which to compute a PTE. Firstly note that we have used the data covariance derived from the marginal likelihood in the definition of $\chi^2$ rather than the simulation estimate $\hat{\mathbfss{C}}$. This is because we wish to account for the broadening of the error bars assigned to the summary statistics coming from imperfect knowledge of the covariance matrix. In the case of a Jeffreys' prior on the true covariance, $\mathbfss{C}_{\mathbfit{y}}$ is given by Equation~\eqref{eq:CyJ}. The null hypothesis is then naturally stated as the assumption that $\mathbfit{y}$ and $\hat{\mathbfss{C}}$ are distributed as $\mathbfit{y} \sim N_{p}(\boldsymbol{\mu}_0,\mathbfss{C}_0)$ and $\hat{\mathbfss{C}} \sim W_{p}[\mathbfss{C}_0/(n-1),n-1]$. This implies that the quantity
\begin{align}
  T_J^2 &\equiv \frac{n-p+1}{(p-1)(n-1)} \left(\mathbfit{y} - \hat{A}\boldsymbol{\mu}_0\right)^\intercal \hat{\mathbfss{C}}^{-1} \left(\mathbfit{y} - \hat{A}\boldsymbol{\mu}_0\right) \nonumber \\
  &= \frac{n-p+1}{(p-1)(n-p-2)}  \left(\mathbfit{y} - \hat{A}\boldsymbol{\mu}_0\right)^\intercal \mathbfss{C}_{\mathbfit{y}}^{J-1} \left(\mathbfit{y} - \hat{A}\boldsymbol{\mu}_0\right) \nonumber \\
  &\equiv \frac{n-p+1}{(p-1)(n-p-2)} \chi^2_J
  \label{eq:T2J}
\end{align}
is distributed as $T^2_J \sim F_{p-1,n-p+1}$, i.e. an $F$-distribution (see e.g.~\citealt{Anderson} for a derivation of this result). Note that in the second line of Equation~\eqref{eq:T2J} we have inserted the Jeffreys' data covariance matrix given in Equation~\eqref{eq:CyJ}, and in the third line defined the quantity $\chi^2_J$. In the limit that $n \rightarrow \infty$ the covariance matrix is known perfectly and $T_J^2$ is just the standard reduced chi-squared test statistic, with the null hypothesis that $(p-1)T_J^2 \sim \chi^2_{p-1}$ (this also follows from the asymptotic properties of the $F$-distribution). Note that the distribution of the test statistic in this case is independent of the unknown quantities $\boldsymbol{\mu}_0$ and $\mathbfss{C}_0$. This is clearly an essential property of test statistics and their assumed distributions under a null hypothesis\footnote{Quantities such as $T_J^2$ are termed \emph{pivotal} or \emph{ancillary} quantities in statistics jargon.}. Once the particular value of $T_J^2$ is computed from our particular realization of $\mathbfit{y}$ and $\hat{\mathbfss{C}}$ we can compute the probability of getting a value at least as large by integrating the $F$-distribution, whose cumulative distribution function is given by a regularized incomplete beta function, for which standard numerical routines exist. We can then test the null hypothesis in the standard way.

How do we generalize the test statistic to include the theoretical model for the covariance matrix $\mathbfss{C}_T$? We wish to retain the property that the test statistic is proportional to $\chi^2$ as defined in Equation~\eqref{eq:chi2}, since we clearly require that this be distributed as $\chi^2_{p-1}$ in the limit that $\mathbfss{C}_T = \mathbfss{C}_0$ and $m \rightarrow \infty$ (or $f_P^{ii} \rightarrow 0$), i.e. when the true covariance is known a priori. We also require that the distribution of the test statistic should be independent of the unknown quantities $\boldsymbol{\mu}_0$ and $\mathbfss{C}_0$. It is tempting to simply define a quantity $T^2$ in analogy with $T_J^2$ in Equation~\eqref{eq:T2J} by replacing $\chi^2_J$ with $\chi^2$ and replacing $n$ with $n+m$ in the prefactor, since this is how the degree-of-freedom parameter of the marginal likelihood is altered when $m \neq 0$. To ensure the correct asymptotic behaviour we could also assume that this test statistic is distributed as $F_{p-1,n+m-p+1}$ under the null hypothesis. However, these definitions would conspire to penalise a prior having $\mathbfss{C}_T = \mathbfss{C}_0$ but finite $m$. This is because the sampling distribution of $T^2$ in this case is not $F_{p-1,n+m-p+1}$, and so the PTEs will not be uniformly distributed, thus systematically penalising this choice of $\mathbfss{C}_T$ (i.e. this particular null hypothesis). We would like to construct a test statistic and null hypothesis which do not penalise the choice $\mathbfss{C}_T = \mathbfss{C}_0$ with finite $m$. In other words, we do not wish for a model-fit to be judged as poor just because the model builder did not have enough confidence in their model when it is in fact correct. Later we will construct tests based on the variances of the model parameters and data covariance matrix which do penalise this underconfidence.

Taking the above considerations into account, the most straightforward choice is to simply use the quantity $\chi^2$ as defined in Equation~\eqref{eq:chi2} as the test statistic, with its distribution under the null hypothesis assumed to be the sampling distribution of $\chi^2$ \emph{when the model covariance} $\mathbfss{C}_T$ \emph{is set equal to the true covariance} $\mathbfss{C}_0$. In Appendix~\ref{app:chi2dist} we prove that this distribution is independent of $\boldsymbol{\mu}_0$ and $\mathbfss{C}_0$. Although we were not able to derive an analytic form for this distribution, we can draw samples from it straightforwardly for each choice of the parameters $(m,n,p)$ and then create a look-up table of PTEs, from which we can interpolate to find the particular PTE of our dataset. The details of this procedure are described in Appendix~\ref{app:chi2dist}. Our PTE is thus defined as
\begin{equation}
  \mathrm{PTE} \equiv \int_{\chi^2}^{\infty} \mathrm{d}x \, p_{\chi^2(\mathbfss{C}_T = \mathbfss{C}_0)}(x).
  \label{eq:PTEdef}
\end{equation}
With this choice of null hypothesis, PTEs computed from $\chi^2$ are uniformly distributed on the interval $[0,1]$ when $\mathbfss{C}_T = \mathbfss{C}_0$, for any choice of $n$ or $m$. We also recover this uniform distribution in the limit of large $n$ for any choice of model covariance. When the model covariance is wrong but the confidence in it is high the PTE will not be uniform, and its measured value may be used to exclude that model given some threshold PTE values. The above construction may be generalized straightforwardly to the case of fitting multiple parameters from the data in the case where the best-fit estimates of the parameters are linear functions of the data.

Finally in this section we note that so far the discussion of assessing the quality of model fits has been focussed on frequentist $\chi^2$-type tests. An alternative approach would be to calculate the Bayesian evidence for each model and then compute posterior odds ratios to discriminate between models, for example different choices of $\mathbfss{C}_T$ or $m$. In our single-parameter model for $\boldsymbol{\mu}$ with a uniform prior on $A$ the Bayesian evidence can be computed analytically, although the final answer is formally ill-defined in the limit that the boundaries of the uniform prior tend to infinity. The evidence ratio can be defined however, and can be converted to a posterior odds ratio assuming equal priors for the competing models. While this approach is useful for comparing different models, it does not offer a way of assessing the quality of a single model, which is why we prefer to work with frequentist methods in this work. In addition, frequentist assessment of model fits will be more familiar to cosmologists, and our approach makes contact with existing methods for inferring parameters from large-scale structure. If one wished to assess model fits in a more Bayesian way, for example with model selection or posterior predictive distributions, our marginal likelihood in Equation~\eqref{eq:marglike} may be used for this.

The marginal likelihood Equation~\eqref{eq:marglike} represents the main result of this work, and the reader uninterested with the details of how to choose the hyperparameter $m$ or the implications for model fitting may now skip to the conclusions in Section~\ref{sec:conc}. In the next few sections we test this likelihood on simulated data and covariance matrices in order to determine the potential reductions in simulation requirements for large-scale structure surveys.

\section{Testing the likelihood on simulations}
\label{sec:sims}

We have constructed a marginal likelihood for the data which depends on a theoretical model covariance $\mathbfss{C}_T$ and a confidence parameter $f_P^{ii}$ (or $m$) dictating the weight this model gets in the data covariance $\mathbfss{C}_{\mathbfit{y}}$. In this section we proceed to investigate the impact of $f_P^{ii}$ and $\mathbfss{C}_T$ on the quality of model fits, parameter variances, and the variance of the error bars we assign to the data through $\mathbfss{C}_{\mathbfit{y}}$. Our ultimate goal is to assess how these choices influence the minimum number of simulations that need to be run in order to attain reasonable errors and model fits, in the hope that we might use a theoretical model covariance to reduce this number.

\subsection{Simulation choices}

As a first step to investigating our data model in the context of surveys of large-scale structure we consider measurements of the real-space matter power spectrum $P(k)$ at $z \approx 0$. We create mock datasets and mock ensembles of simulations for $\hat{\mathbfss{C}}$ by generating a large number ($4 \times 10^5$) of Gaussian realizations (to be consistent with the assumptions of Section~\ref{sec:theory}) of $P(k)$ in $p = 21$ bins in the wavenumber $k$. The mean and covariance matrix of these realizations were set equal to the mean and covariance of 719 measurements of $P(k)$ from independent N-body simulation snapshots at $z=0.042$ from the SLICS\footnote{\url{http://slics.roe.ac.uk}} simulation suite~\citep{2015MNRAS.450.2857H,2018arXiv180504511H}. These were measured on a fine grid in $k$ and then rebinned by taking an average of $P(k)$ within each of our $p$ bins weighted by the number of $\mathbf{k}$-modes in each bin. The N-body simulations were run in a $(505 \, h^{-1} \, \mathrm{Mpc})^3$ cubic box with a WMAP9+BAO+SN cosmology~\citep{2013ApJS..208...19H} having flat $\Lambda$CDM parameters $(\Omega_m, \Omega_{\Lambda}, \Omega_b, \sigma_8, h, n_s) = (0.2905,0.7095,0.0473,0.831,0.6898,0.969)$. The $P(k)$ measurements in our $p$ wavenumber bins contain negligible shot noise, and are robust to changes in the simulation resolution. Note that the details of the simulations are unimportant here, as we only require a representative mean and covariance matrix. We chose our $k$-bins to be linearly spaced in $k$ between roughly $0.035 \, h \, \mathrm{Mpc}^{-1}$ and $1 \, h \, \mathrm{Mpc}^{-1}$ such that we cover the linear and non-linear regimes. In Section~\ref{subsec:dim} we consider the cases $p=11$ and $p=31$ to study the impact of dimensionality on our results.

\begin{figure}
  \includegraphics[width=\columnwidth]{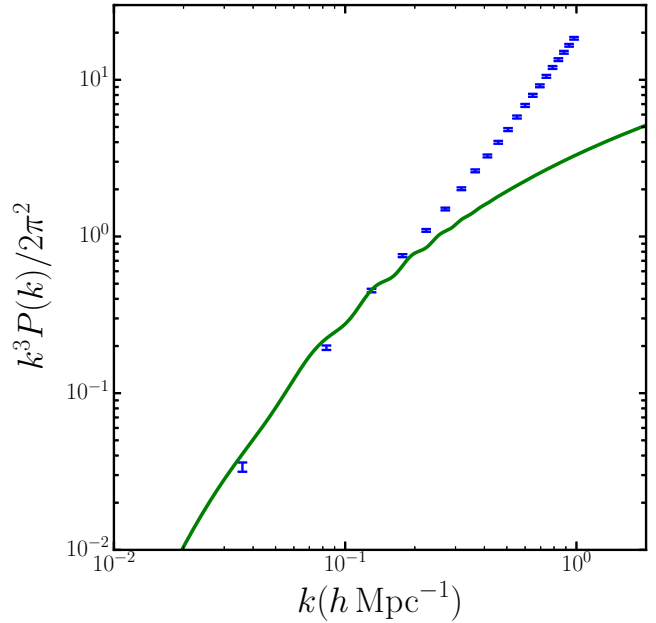}
  \caption{Mean dimensionless power spectrum $k^3P(k)/2\pi^2$ of the simulations (blue points), with errors computed from the empirical variance of the simulations. The green solid line is the linear prediction for this redshift ($z=0.042$).}
  \label{fig:pk}
\end{figure}

In Figure~\ref{fig:pk} we plot the mean of the rebinned dimensionless power spectrum estimates from the N-body simulations, along with the linear theory prediction at this redshift, with error bars given by the empirical variance from the simulations. This figure demonstrates that most of our $k$-bins probe the non-linear regime of structure formation, with only the largest three or four scales accurately modelled with linear theory.

\begin{figure}
  \includegraphics[width=\columnwidth]{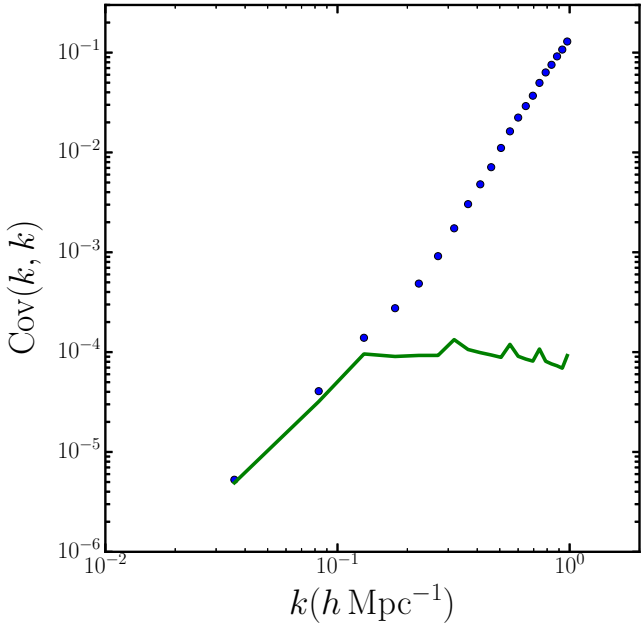}
  \caption{Diagonal elements of the covariance matrix of the dimensionless power spectra estimated from all the simulations (blue points), and the linear prediction for this redshift (green curve, $z=0.042$). The step-like features in the linear prediction arise from jumps in the number of modes contributing to each bin.}
  \label{fig:covd}
\end{figure}

In Figure~\ref{fig:covd} we plot the diagonal elements of the covariance matrix from our simulations, along with the linear prediction given by
\begin{equation}
  \mathrm{Cov}_L(k_1,k_2) = 2\frac{\left[k_1^3P_L(k_1)/2\pi^2\right]^2}{N(k_1)} \delta_{k_1,k_2},
  \label{eq:covL}
\end{equation}
where $N(k_1)$ is the number of modes contributing to the bin centred on $k_1$, and $\delta_{k_1,k_2}$ is the Kronecker delta. The linear power spectrum is computed for our cosmology using CAMB~\citep{2000ApJ...538..473L, 2012JCAP...04..027H}. Since the relevant quantity for us is the rebinned power spectrum, we average the value of $P(k)^2$ over the bin centred on $k_1$ weighted by the number of modes in each of the finer $k$-bins used in the SLICS power spectrum estimates. Note that the total number of $\mathbf{k}$-modes contributing to our bins varies between each bin, which can give rise to step-like features in the linear covariance, as seen in the green curve in Figure~\ref{fig:covd}. This figure demonstrates that the linear prediction is only good for the first two or three $k$-bins at $z=0.042$, i.e. linear theory breaks down at slightly large scales than for the matter power spectrum itself, c.f. Figure~\ref{fig:pk}.

\begin{figure}
  \includegraphics[width=\columnwidth]{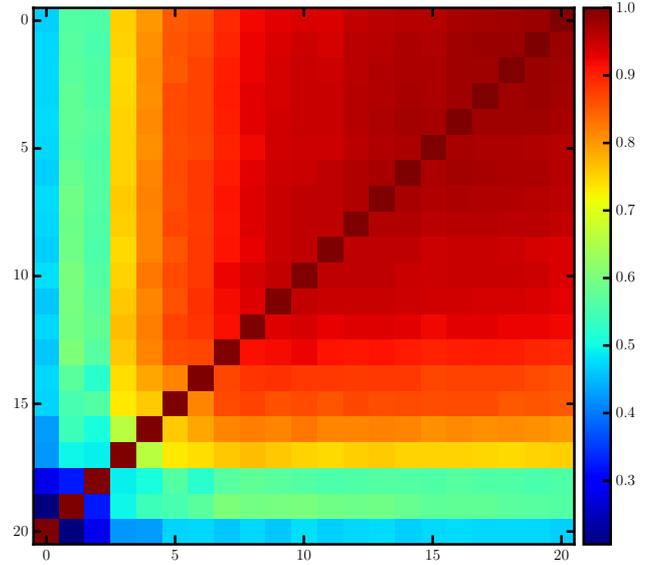}
  \caption{Correlation matrix estimated from all the simulations, for the $k$-bins shown in Fig.~\ref{fig:covd}. Low $k$-bins (large scales) are in the bottom-left, high $k$-bins (small scales) are in the top-right. The value of the dimensionless correlation coefficient for each bin is given by the colour bar.}
  \label{fig:corr}
\end{figure}

In Figure~\ref{fig:corr} we plot the correlation matrix of our power spectrum estimates. Equation~\eqref{eq:covL} tells us that this matrix should be diagonal, which is clearly only a reasonable approximation for the largest three $k$-bins, consistent with the diagonal elements shown in Figure~\ref{fig:covd}.

\subsection{Partitioning the simulations}
\label{subsec:partition}

Our ultimate goal is to investigate how the minimum number of simulations that need to be run to form $\hat{\mathbfss{C}}$ changes for different choices of the model $\mathbfss{C}_T$ and our confidence in that model $f_P^{ii}$. To this end, we partition our set of $4\times 10^5$ simulated $P(k)$ measurements into mock sets of `data' and `simulations', each set corresponding to a different choice of $n$, the number of simulations going into the covariance matrix estimate. We choose 30 values of $n$ between $n=2$ and $n=205$, with a spacing of $\Delta n = 7$. The upper limit here is based on the one-parameter model considered in Section~\ref{subsec:post}. For this model, a value of $n=200$ ensures a precision of 10\% on the diagonal elements of the estimated covariance matrix (i.e. $f_{\hat{\mathbfss{C}}}^{ii} = 0.1$), and a 1\% degradation of the average Fisher variance of $A$ (independent of $p$) from marginalizing over the covariance matrix with a Jeffreys' prior, see Equation (42) of \citet{2017MNRAS.464.4658S}. The same value also gives a $\sim 1\%$ increase in the true average posterior variance of $A$ assuming a Jeffreys' prior. In Appendix~\ref{app:varAJapprox} we derive an accurate approximation for this quantity. A value of $n=200$ is thus clearly sufficient for this inference problem, and so we cap the maximum value of $n$ at roughly this value.

For each $n$ we partition our $P(k)$ measurements into `data' and `simulation ensembles'. The data consists of $n_{\mathrm{obs}} = 400$ sets of $p=21$ vectors for each $n$, while the simulation ensembles are used to form $n_{\mathrm{ens}}(n)$ realizations of $\hat{\mathbfss{C}}$ for each $n$. We will ultimately estimate averages of quantities such as the PTE over the mock data and mock simulation ensembles for each $n$, so we need to make $n_{\mathrm{obs}}$ and $n_{\mathrm{ens}}(n)$ large enough for these averages to converge, yet small enough that the computations are not too expensive. Since quantities such as $\mathbfss{C}_{\mathbfit{y}}$ are noisy at low values of $n$ we impose that  $n_{\mathrm{ens}}(n) \propto n^{-1}$. This typically leads to $n_{\mathrm{ens}} \approx 6500$ for $n=2$ and $n_{\mathrm{ens}} \approx 60$ for $n=205$. The sum $\sum_n \left[ n \times n_{\mathrm{ens}}(n) + n_{\mathrm{obs}}\right] = 4 \times 10^5$, i.e. the total number of Gaussian simulations we have generated from the SLICS mean and covariance. We confirmed that our results were stable to increasing $n_{\mathrm{obs}}$ and $n_{\mathrm{ens}}(n)$.

\subsection{Theoretical model covariance choices}
\label{subsec:CTs}

We wish to study the impact of different choices for $\mathbfss{C}_T$ and $f_P^{ii}$ on the final model fits. State-of-the art models for the covariance matrix of large-scale structure two-point statistics include the halo model~\citep{2000MNRAS.318..203S, 2000MNRAS.318.1144P, 2002PhR...372....1C} and perturbation theory-based models~\citep{2002PhR...367....1B,2016PhRvD..93l3505B, 2017MNRAS.466..780M, 2017JCAP...11..051B, 2018arXiv180704266B}. As we are only aiming for a proof-of-concept here we will not consider these approaches but instead study a simplified set of models which capture some of the key features. We choose six models for $\mathbfss{C}_T$, summarized below.

\subsubsection*{The true covariance matrix $\mathbfss{C}_0$}

This choice simply sets $\mathbfss{C}_T = \mathbfss{C}_0$. While this is clearly not a realistic model (we do not know $\mathbfss{C}_0$), it serves as a valuable sanity check on some of the results. In particular, since the null hypothesis is that $\mathbfss{C}_T = \mathbfss{C}_0$, we expect the PTE computed from each data-simulation-ensemble pair to be uniformly distributed, and hence $\langle \mathrm{PTE} \rangle = 0.5$ when averaging over all data and simulation ensembles for each $n$.

\subsubsection*{The true covariance matrix increased by 10\%}

This choice sets $\mathbfss{C}_T = 1.1 \times \mathbfss{C}_0$. While again this is not a realistic choice, it will allow us to gauge the impact of our theory overestimating each element of the covariance matrix.

\subsubsection*{The true covariance matrix with diagonals increased by 10\%}

This choice scales all the diagonal elements of $\mathbfss{C}_0$ by a factor 1.1, i.e. preserving the off-diagonal elements but increasing the diagonal elements by 10\%. The correlation matrix is thus decreased by roughly 10\% as well.

\subsubsection*{The true covariance matrix with off-diagonals decreased by 10\%}

As a complementary choice to scaling the diagonal elements, this choice scales all the off-diagonal elements of $\mathbfss{C}_0$ by a factor 0.9, leaving the diagonal elements unchanged. This reduces the correlation matrix by 10\%. Note that we do not consider the case of overestimation of the off-diagonal elements since this would require carefully ensuring that the covariance is still positive definite.

\subsubsection*{Diagonal elements of $\mathbfss{C}_0$}

This choice sets all the off-diagonal elements of $\mathbfss{C}_0$ to zero but keeps its diagonal elements. The motivation behind this choice is to investigate the impact of ignoring the off-diagonal elements completely, in contrast to the case where we merely consider a small underestimation of these elements.

\subsubsection*{Linear covariance matrix $\mathbfss{C}_L$}

This choice just uses the linear prediction from Equation~\eqref{eq:covL}. As we see from Figure~\ref{fig:covd} and Figure~\ref{fig:corr}, this model underestimates the true diagonal elements and ignores the off-diagonal elements of the covariance matrix.

\subsection{Prior width choices}

As well as investigating the impact of different choices of $\mathbfss{C}_T$ on model fits and parameter inference, we also need to investigate the impact of the prior width, specified through $f_P^{ii}$ (the standard deviation of the diagonal elements in units of the mean for the prior, see Equation~\eqref{eq:fpdef}) or $m$ (the degree-of-freedom parameter of the prior). The precise level of confidence will depend on a combination of how many simulations were available to test the model against, how biased the model appears to be with respect to those simulations, and whether or not the model can predict its own accuracy (as in effective field theory or response function approaches), see the discussion in Section~\ref{subsec:priors}.

We consider seven values of $f_P^{ii}$, with corresponding values for $m$ (assuming $p=21$) given by $(f_P^{ii},m)$ = $(0.01,20024)$, $(0.05,824)$, $(0.10,224)$, $(0.20, 74)$, $(0.35,40.3)$, $(0.50, 32)$, and $(\infty,24)$. We remind the reader again here that small values of $f_P^{ii}$ correspond to high confidence in the model and low values to low confidence. We also remind the reader that the off-diagonal elements have necessarily broader priors, see Equation~\eqref{eq:fpij}.

Unlike the different choices of $n$ we use the same data and simulations for each choice of $f_P^{ii}$. We also ran a model having $(f_P^{ii},m) = (0.00,\infty)$, but found this to be indistinguishable from $(f_P^{ii},m) = (0.01,20024)$, suggesting that 1\% confidence in the diagonal elements is roughly equivalent to complete confidence in these elements.

\section{Results}
\label{sec:results}

For each value of $n$ and $f_P^{ii}$ we computed the average value of the posterior variance of $A$, defined in Equation~\eqref{eq:Astats}, to assess how this particular parameter variance is affected by prior model choices. We also computed the average PTE as defined in Equation~\eqref{eq:PTEdef} across the simulations, in order to gauge the expected impact of different choices of $\mathbfss{C}_T$ and $f_P^{ii}$ on the quality of model fits - too much confidence in a model covariance which is wrong will result in misestimated error bars for the data and hence an unacceptable PTE even when the template $\boldsymbol{\mu}_0$ is the truth. In this section we present the results of these tests, along with an assessment of how the variance in $\mathbfss{C}_{\mathbfit{y}}$ itself changes as we change $f_P^{ii}$. We will then combine these tests to determine the minimum number of simulations that need to be run to achieve acceptable values of the PTE, the parameter variance, and the data covariance variance, for the different models and $f_P^{ii}$ values. Finally we will test the sensitivity of our results to the dimensionality of the data vector $p$.

\subsection{Parameter variance tests}
\label{subsec:varAtests}

\begin{figure*}
  \includegraphics[width=2\columnwidth]{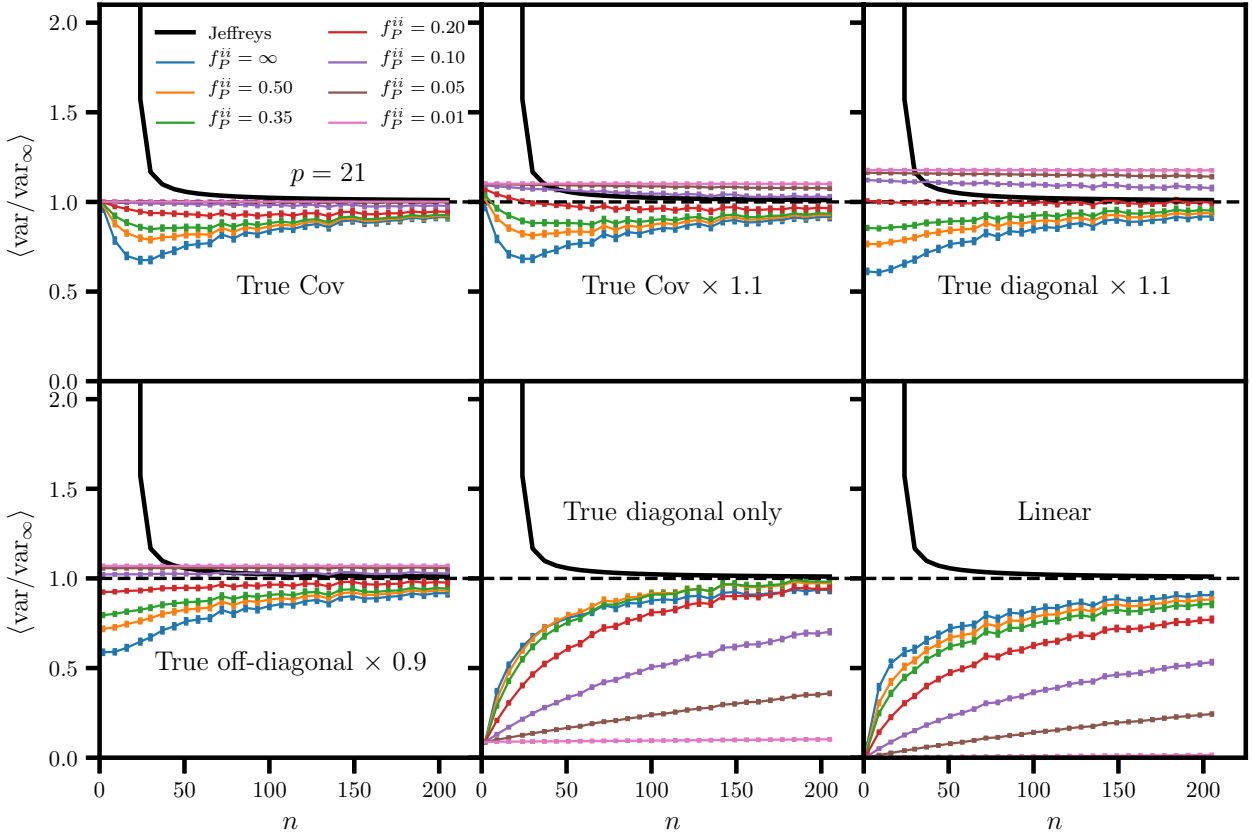}
  \caption{Average variance of the amplitude $A$ relative to its $n=\infty$ limit when the theory covariance is set to the true covariance (top-left panel), the true covariance scaled by 1.1 (top-middle panel), the true covariance with its diagonal elements scaled by 1.1 (top-right panel), the true covariance with its off-diagonal elements scaled by 0.9 (bottom-left panel), the true covariance with its off-diagonal elements set to zero (bottom-middle panel), and the linear covariance (bottom-right panel). In each panel we plot the dependence on the number of simulations $n$ for different choices of the prior width $f_P^{ii}$, for $f_P^{ii} = \infty$ (blue), 0.50 (orange), 0.35 (green),  0.20 (red), 0.10 (purple), 0.05 (brown), and 0.01 (pink). Also plotted is the variance when the Jeffreys' prior is adopted (black solid curve), which diverges when $n=p+1$. Curves are ordered bottom-to-top in each panel for the highest-to-lowest values of $f_P^{ii}$ except for the bottom-middle and bottom-right panels where the trend is reversed. The dashed horizontal line denotes a value of unity.}
  \label{fig:varratplot}
\end{figure*}

How does placing increasing confidence in a model covariance matrix impact parameter constraints? In Figure~\ref{fig:varratplot} we plot the average posterior variance on $A$ relative to its $n=\infty$ limit (given by $1/\boldsymbol{\mu}_0 \mathbfss{C}_0^{-1}\boldsymbol{\mu}_0$), again for different choices of the prior parameters, and for $p=21$. We also plot the equivalent value for the Jeffreys' prior using the (very accurate) approximation of Equation~\eqref{eq:varAJapprox}, which diverges when $n=p+1$ as the data covariance matrix becomes singular at this point. This curve is independent of the model, and increases from unity as $n$ is lowered due to the extra variance incurred as a result of marginalizing over the covariance matrix~\citep{2016MNRAS.456L.132S, 2017MNRAS.464.4658S}.

When the model covariance is set to the true covariance, it is somewhat unclear how increasing the confidence in the model should impact the variance of the amplitude parameter $A$. As confidence in the model become very high we should recover the $n=\infty$ limit, as seen in the top-left panel of Figure~\ref{fig:varratplot}. However, as we decrease this confidence the parameter variance becomes monotonically \emph{lower} than this value. In other words, upon using a less informative prior the average parameter variance actually decreases rather than increases. The reason for this is that the information gained by reducing $f_P^{ii}$ is mainly going into making $\mathbfss{C}_{\mathbfit{y}}$ less noisy, whereas the mean of $\mathbfss{C}_{\mathbfit{y}}$ is unchanged and is simply equal to $\mathbfss{C}_0$, independent of  $f_P^{ii}$. The average posterior variance of $A$ is primarily sensitive to the average of $\mathbfss{C}_{\mathbfit{y}}$, so the dependence on $f_P^{ii}$ is generally quite weak. The primary impact of changing $f_P^{ii}$ is then through its impact on random fluctuations in $\mathbfss{C}_{\mathbfit{y}}$ which propagate to fluctuations in $\mathrm{var}(A)$. We find that the term in square brackets in Equation~\eqref{eq:Astats} is subdominant and $\mathrm{var}(A) \sim 1/\boldsymbol{\mu}_0 \mathbfss{C}_{\mathbfit{y}}^{-1} \boldsymbol{\mu}_0$. At low values of $n$ and high values of $f_P^{ii}$ the data covariance $\mathbfss{C}_{\mathbfit{y}}$ is noisy, and $1/\boldsymbol{\mu}_0 \mathbfss{C}_{\mathbfit{y}}^{-1} \boldsymbol{\mu}_0$ is biased low\footnote{This is to be compared with the exact average of $\left(\boldsymbol{\mu}_0 \hat{\mathbfss{C}}^{-1} \boldsymbol{\mu}_0 \right)^{-1}$ over its (Gamma) sampling distribution, which is $\frac{n-p}{n-1}\left(\boldsymbol{\mu}_0 \mathbfss{C}_0^{-1}\boldsymbol{\mu}_0\right)^{-1}$, i.e. biased low.}, and so the variance of $A$ is biased low, as seen in the top-left panel of Figure~\ref{fig:varratplot}. The errors on the parameters are also quite non-Gaussian in this regime, and so the error bars presented in this regime should be treated with caution. Thus, one should be very cautious about making judgements when the data covariance matrix is noisy; later on we will impose thresholds on this noise such that low values of $n$ and $m$ are ruled inadmissible.

When the true covariance is scaled by 10\% the variance on $A$ is also increased by roughly 10\% (top-middle panel), since the dominant part of $\mathrm{var}(A)$ is proportional to $\mathbfss{C}_{\mathbfit{y}}$. If instead we increase only the diagonal elements by 10\% there is an extra boost to this parameter variance over scaling all the elements, with similar behaviour seen when reducing the off-diagonal elements, see the bottom-left panel of Figure~\ref{fig:varratplot}. In contrast, when the off-diagonal elements are ignored completely the variance is systematically \emph{lower} compared to the $n=\infty$ limit.

\begin{figure*}
  \includegraphics[width=2\columnwidth]{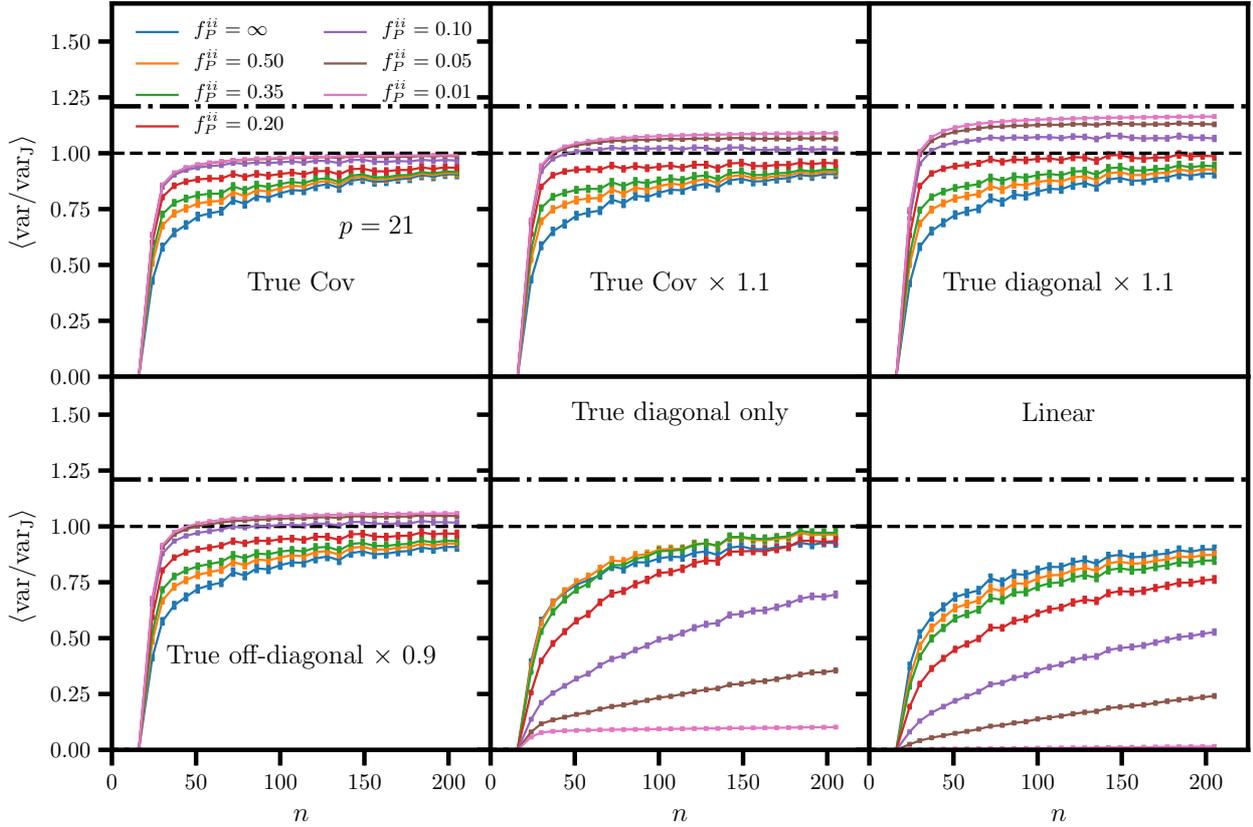}
  \caption{Average variance of the amplitude $A$ relative to its Jeffreys'-prior value for different choices of the theory covariance and number of simulations $n$. Panels and curves denote the same quantities as in Fig.~\ref{fig:varratplot}. Curves are ordered bottom-to-top in each panel for the highest-to-lowest values of $f_P^{ii}$ except for the bottom-middle and bottom-right panels where the trend is reversed. The dashed horizontal line denotes a value of unity, the dot-dashed line denotes 10\% extra error (i.e. square-root of the variance) in $A$ relative to the Jeffreys' value.}
  \label{fig:dvaratplot}
\end{figure*}

To understand this behaviour, consider for example fitting a straight line to two Gaussian data points having conditional errors $\sigma_1$ and $\sigma_2$ and correlation $\rho$. The variance on the amplitude is $(1-\rho^2)/(X_1^2 + X_2^2 - 2X_1X_2\rho)$, where $X_i = \mu_i/\sigma_i$ is the conditional signal-to-noise on data point $i$ with $\mu_i$ the model prediction. Differentiating with respect to $\rho$ and setting $X_1X_2 > 0$, we see that $\mathrm{var}(A)$ increases with $\rho$ whenever $\rho < \mathrm{min}(X_1/X_2,X_2/X_1)$, and decreases with $\rho$ whenever $\mathrm{min}(X_1/X_2,X_2/X_1) < \rho < 1$, with the opposite behaviour if $X_1X_2 < 0$. In the case of complete correlation $\rho=1$ the variance is zero (perhaps counterintuitively) since there are two data points for two unknowns; $A$ and the common noise between the two points. Thus the parameter variance can either decrease or increase upon changing $\rho$ depending on its starting value. In our case we have $X_1X_2 > 0$ since $P(k)>0$, and the data points are strongly correlated, see Figure~\ref{fig:corr}. Since the conditional signal-to-noise on the power spectrum estimates varies strongly across different scales this puts us in the regime where small (e.g. 10\%) decreases in $\rho$ increase $\mathrm{var}(A)$, explaining the behaviour in the middle panels of Figure~\ref{fig:varratplot}. If instead $\rho$ is set to zero the variance decreases, since then the second data point contains independent information from the first data point which conspires to reduce the variance for the particular correlation matrix of $\mathbfss{C}_T$. We believe this simplified example captures the main effects in Figure~\ref{fig:varratplot}, and also demonstrates that correlations between data points have a non-trivial effect on parameter variances. Note that the results at low values of $n$ and $m$ are biased by fluctuations in the data covariance matrix, as discussed above.

When the linear covariance matrix is assumed we see that high confidence in the prior leads to very low parameter variance. This makes intuitive sense since the diagonal elements are severely underestimated (see Figure~\ref{fig:covd}), and the correlations are neglected.

In Figure~\ref{fig:dvaratplot} we again plot the amplitude variance but this time divided by its Jeffreys' prior value. These ratios thus go to zero when $n \leq p+1$. This quantity is of interest, since it tells us the relative increase or decrease of the parameter variance at finite $n$ when marginalizing against our informative prior. We also plot a 10\% increase threshold on the error (i.e. square-root of the variance) as the dot-dashed line on this figure. We see that the increase in parameter error is never greater than 10\% for any of our chosen models, with the largest increase coming when we have high confidence in a model where the diagonal elements have been overestimated by 10\%. In most cases the variance is lower than Jeffreys', for the reasons discussed above.

\begin{figure*}
  \includegraphics[width=2\columnwidth]{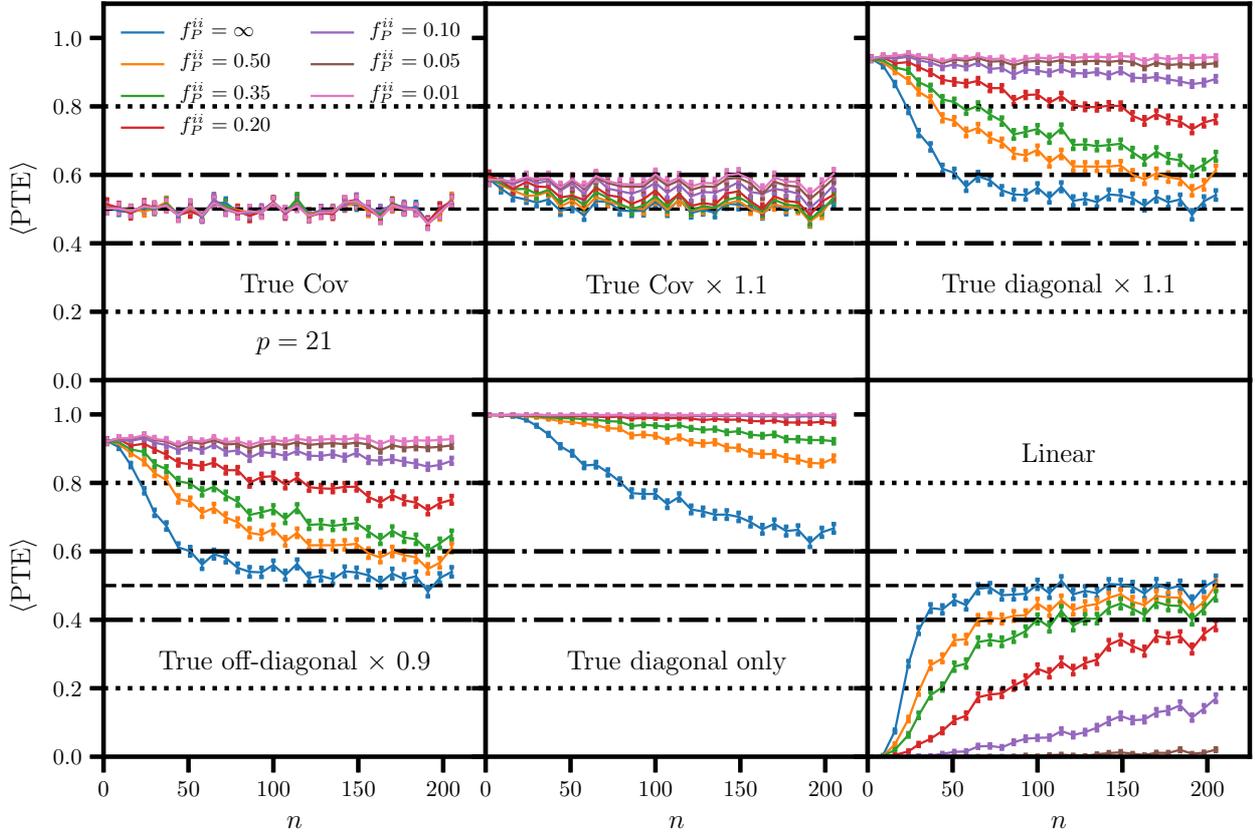}
  \caption{Average PTE for different choices of the theory covariance and number of simulations $n$.  Panels and curves denote the same quantities as in Fig.~\ref{fig:varratplot}. Curves are ordered bottom-to-top in each panel for the highest-to-lowest values of $f_P^{ii}$ except for the bottom-right panel where the trend is reversed. The dashed line shows the $\langle \mathrm{PTE} \rangle = 0.5$ value, the dot-dashed lines bound the $0.5 \pm 0.1$ thresholds, and the dotted lines bound the $0.5 \pm 0.3$ thresholds.}
  \label{fig:PTEplot}
\end{figure*}

In the case that the covariance is overestimated elementwise by 10\% (top-middle panel of Figure~\ref{fig:dvaratplot}), the choice of $f_P^{ii}$ which gives a parameter variance closest to the Jeffreys' result is $f_P^{ii} \approx 0.10$ (purple curve). For this particular scenario the prior width on the diagonal elements is the same as the bias in the model, such that increase in parameter variance coming from the overestimated theory covariance is almost exactly balanced by the decrease due to the uncertainty in the true covariance matrix given this model.

We close this section with the caveat that use of a single amplitude parameter to construct parameter variance tests is clearly limited. Poor covariance models could have large effects on the posterior distributions of parameters which impart scale-dependent effects in the power spectrum, while leaving the amplitude variance unchanged or reduced. This is potentially true of parameters whose impact on the power spectrum is greatest where the covariance model is poor, for example on non-linear scales where realistic models are likely to be least accurate. However, as long as one increases the prior width to account for the inaccuracy, this should not give parameter biases greater than the $1 \sigma$ posterior width but should just broaden the final posterior. A thorough investigation would involve properly studying the sensitivity of the full posterior to the prior covariance choices. This is worthy of further study, which we defer to a future work.

\subsection{PTE tests}
\label{subsec:PTEtests}

We have seen that parameter variances can be either reduced or increased when a theoretical covariance matrix is folded into the inference procedure. Clearly we would like to reduce the variance, but not at the expense of making our error bars artificially small. A metric for assessing the `quality' of the error bars is provided by the PTE, introduced in Section~\ref{subsec:PTEs}. A low value of the PTE implies a high value of $\chi^2$ for the best-fitting parameters; in other words, the probability of obtaining a value of $\chi^2$ at least as big as the observed value is low even when the null hypothesis is `true', i.e. when it matches the sampling distribution that produced the observations. Since many other systematic errors can reduce the quality-of-fit of the best-fitting model, it is important to ensure that a mis-specified covariance matrix does not dominate the PTE budget.

In Figure~\ref{fig:PTEplot} we plot the average PTE as a function of $n$ for each of the six covariance models listed in Section~\ref{subsec:CTs}, and for different choices of $f_P^{ii}$, for $p=21$. The error bars on these plots were estimated from the empirical scatter of the simulations, and are comparable to the scatter in the measurements assuming a smooth model for the points. Note that the points are independent for different $n$ at fixed $(\mathbfss{C}_T,f_P^{ii})$ but not for different $(\mathbfss{C}_T,f_P^{ii})$ at fixed $n$. The black dashed line shows the value 0.5, which should be the mean by construction if $\mathbfss{C}_T = \mathbfss{C}_0$, and the black dot-dashed and dotted lines show PTE thresholds of $0.5 \pm 0.1$ and $0.5 \pm 0.3$ respectively.

The top-left panel of Figure~\ref{fig:PTEplot} serves as a sanity check, as we should have $\langle \mathrm{PTE} \rangle = 0.5$ by construction, which is indeed satisfied to within the error bars for each $n$ and $f_P^{ii}$. When we scale up the covariance model by 10\% (top-middle panel), the error bars on the data are all too large and hence the PTE tends to be greater than 0.5, with this effect becoming more severe as confidence in the model is increased. For the case of $f_P^{ii} = 0.01$ the theory model overwhelms the simulation estimate $\hat{\mathbfss{C}}$ at all the values of $n$ we consider, and the average PTE is just below 0.6, i.e. the best fit model is still a reasonable fit to the data. At sufficiently high values of $n$ the estimate $\hat{\mathbfss{C}}$ becomes more important, and we revert back to $\langle \mathrm{PTE} \rangle = 0.5$ since the null hypothesis becomes closer to the true sampling distribution. This transition happens at lower $n$ for larger values of $f_P^{ii}$ (lower confidence in the model), with the blue points for example transitioning from 0.6 to 0.5 by $n \approx 50$. At the lowest values of $n$ the weight is all on the theory, and so all the curves asymptote to each other - this is true for every model choice.

We see similar but more extreme behaviour when only the diagonal elements are increased by 10\%. When confidence in this model is high, the PTE tends to be $\gtrsim 0.9$, and the same behaviour is seen when we reduce the off-diagonal elements by 10\%. The main effect here is the 10\% reduction in the correlation matrix, common to both these models. Highly correlated data points exhibit very little scatter with respect to a best-fitting model. If we neglect these correlations our $\chi^2$ looks too low, and the PTE is too high. This is what we see in the top-right and bottom-left panels of Figure~\ref{fig:PTEplot}; when the model covariance exhibits correlations which are lower than the truth, the PTE is pushed high. Thus it appears that getting the correlations correct will be very important for getting good model fits from planned large-scale structure surveys. Considering that the true correlations are already high for most of the $k$-bins (Figure~\ref{fig:corr}), this implies that the PTE is highly sensitive to small changes in the elements of the correlation matrix for a survey probing non-linear scales.

We see this again in a more extreme fashion when the off-diagonal elements are neglected all together (bottom-middle panel of Figure~\ref{fig:PTEplot}). Sensitivity to the confidence in the model is now very high, with even very broad priors still being sufficiently informative that the model fits are very poor. For this model, an acceptable PTE is only reached for either a very large number ($n \gg 200$) of simulations or a very uninformative prior; even in the latter case we have a degradation in the model fit for $n \lesssim 200$, suggesting that this model is sufficiently poor that it should not be included in the likelihood at all, and a Jeffreys' prior should instead be used. The same can be said of the linear covariance model (bottom-right panel), where now the effect of an underestimated correlation matrix is actually overwhelmed by the underestimated diagonal elements (see Figure~\ref{fig:covd}). The low diagonal elements push the PTE low (higher $\chi^2$), in opposition to the influence of the low off-diagonal elements. The net effect is that reasonable PTEs can be obtained even for $n \approx 75$ if the prior is sufficiently broad. As we shall see later though, low weight on the prior combined with small numbers of simulations give a noisy data covariance matrix, which is clearly undesirable.

In the previous section we saw that when the theory covariance matrix is biased high by 10\%, comparable parameter variances to the Jeffreys' case can be achieved by tuning the prior width on the diagonal elements to be equal to this bias. The purple curve in the top-middle panel of Figure~\ref{fig:PTEplot} suggests that this choice of $f_P^{ii}$ does not give an optimal PTE, with better model fits available for broader priors. Setting $f_P^{ii} = 0.10$ still gives a reasonable PTE however, suggesting that overall this choice gives performance quite similar to the Jeffreys' prior case, but requiring much fewer simulations - exactly how much fewer will depend on the thresholds one places on changes to parameter variances and PTEs.

The average PTEs thus offer insight into the sensitivity of the quality of model fits to the different covariance models and our confidence in these models. That we can study these effects as a function of confidence rather than the binary choice of simulation-only or theory-only is a great advantage of our approach, and tests similar to these could be conducted with more realistic models for the covariance matrix.

We note finally that the deviation of the average PTE from 0.5 is only a rough measure of the quality of the model fit. A more informative measure would be to study the full sampling distribution of the PTE and compare it against the uniform distribution - the cumulative probability version of this test is known as a quantile-quantile plot, and has recently been adopted in photometric redshift methodology~\citep{2016MNRAS.457.4005W}. Applying this diagnostic to covariance matrix estimation is an interesting line of research which we defer to a future work.

\subsection{Data covariance variance tests}
\label{subsec:varCytests}

As well as demanding reasonable parameter variance and PTEs when the model $\boldsymbol{\mu}_0$ is the truth, it is clearly desirable that the error bars we assign to the data through $\mathbfss{C}_{\mathbfit{y}}$ should not be too noisy. Moreover, we have seen that noisy data covariances lead to average parameter variances which appear abnormally low due to the propagation of this noise, and we do not wish to draw overoptimistic conclusions from the data based on a chance fluctuation in $\hat{\mathbfss{C}}$ or make the parameter posterior too non-Gaussian.

The variance of the data covariance is given in Equation~\eqref{eq:varCy}. The most intuitive and simple measure of noise in $\mathbfss{C}_{\mathbfit{y}}$ would be the standard deviation of the diagonal elements in units of the mean, given by
\begin{equation}
  \frac{\sqrt{\mathrm{var}(\mathbfss{C}_{\mathbfit{y},ii})}}{\langle \mathbfss{C}_{\mathbfit{y},ii} \rangle } = \frac{\sqrt{2(n-1)}\mathbfss{C}_{0,ii}}{(n-1)\mathbfss{C}_{0,ii} + (m-p-1)\mathbfss{C}_{T,ii}}.
\end{equation}
This metric depends on the unknown covariance matrix $\mathbfss{C}_{0}$ and hence could not be used in a realistic setting. The quantity that \emph{is} independent of $\mathbfss{C}_{0}$ is the standard deviation of the diagonal elements in units of the true value, given by
\begin{equation}
  f_y^{ii} \equiv  \frac{\sqrt{\mathrm{var}(\mathbfss{C}_{\mathbfit{y},ii})}}{\mathbfss{C}_{0,ii}} = \frac{\sqrt{2(n-1)}}{n+m-p-2}.
  \label{eq:fydef}
\end{equation}
These two quantities are roughly equal when $\mathbfss{C}_{T,ii} \approx \mathbfss{C}_{0,ii}$, as is the case for all our models except the linear model. Since the linear data covariance is significantly biased away from the truth on most scales, Equation~\eqref{eq:fydef} provides only an approximate measure of the width of the covariance. However, its independence from unknown quantities makes $f_y^{ii}$ a useful metric for assessing the variance in $\mathbfss{C}_{\mathbfit{y}}$.

\begin{figure}
  \includegraphics[width=\columnwidth]{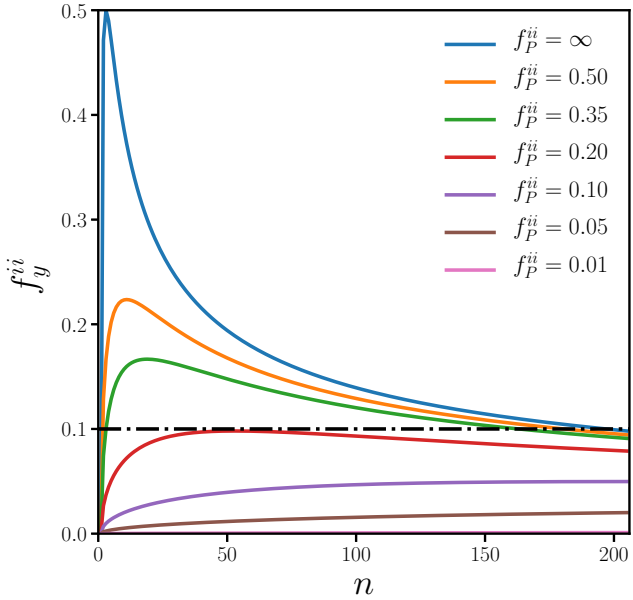}
  \caption{Standard deviation of the diagonal elements of the data covariance matrix (derived from the likelihood) in units of the square-root of the corresponding diagonal element of the true covariance matrix, as a function of the number of simulations $n$. We show different choices of the prior width $f_P^{ii}$, for $f_P^{ii} = \infty$ (blue, highest curve), 0.50 (orange, second-highest curve), 0.35 (green, third-highest curve),  0.20 (red, fourth-highest curve), 0.10 (purple, fifth-highest curve), 0.05 (brown, sixth-highest curve), and 0.01 (pink, lowest curve) The dot-dashed line denotes a 10\% threshold.}
  \label{fig:covcy}
\end{figure}

In Figure~\ref{fig:covcy} we plot the quantity $f_y^{ii}$ as a function of $n$ and $f_P^{ii}$, which is independent of $p$. As expected, the variance on $\mathbfss{C}_{\mathbfit{y}}$ goes to zero when either $n$ becomes large or $f_P^{ii}$ becomes small, due to the suppression of noise in the simulation estimate $\hat{\mathbfss{C}}$ and shrinkage to the noise-free model respectively. There is a maximum when $n = m - p$, i.e. when there is equal weight in the data covariance between simulations and model. When $n \gg m,p$ we have $f_y^{ii} \rightarrow \sqrt{2/n}$ and all curves asymptote to each other, recovering the standard Gaussian result (see, e.g. \citealt{2011ApJ...726....7T}).

We have also plotted a 10\% threshold on Figure~\ref{fig:covcy}. This was chosen to ensure reasonable noise properties for the data covariance matrix, and is at roughly the level demanded of simulated covariances by recent large-scale structure surveys that have made use of $\hat{\mathbfss{C}}$ (e.g. \citealt{2012MNRAS.426.1262H, 2013MNRAS.430.2200K}). This threshold excludes the regime  where the number of simulations is small and the confidence in the prior is low. Note that we never consider the value $n=1$ in these tests.

\subsection{Minimum number of simulations}
\label{subsec:minn}

We have seen how different choices of the prior width $f_P^{ii}$ for each choice of $\mathbfss{C}_T$ impact upon the model fit quality, the amplitude variance, and the variance of the data covariance matrix. By demanding that each of these quantities is not degraded too much by our choices of $f_P^{ii}$ and $n$, we can determine the minimum number of simulations $n_{\mathrm{min}}$ which need to be run for each choice of the model covariance matrix.

\captionsetup[subfigure]{labelformat=empty}
\begin{figure*}
  \centering
  \subfloat[]{\includegraphics[width=\columnwidth]{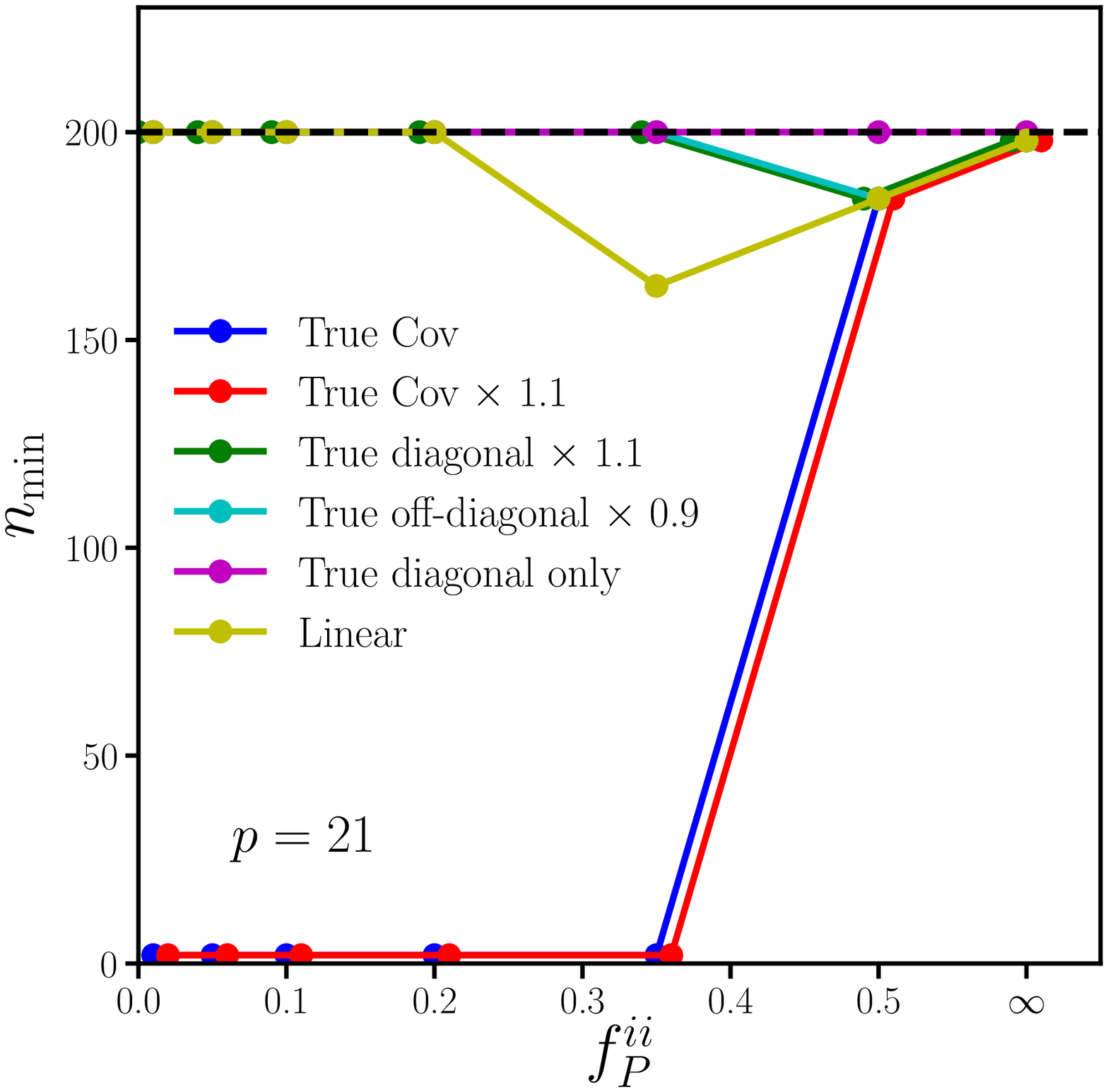}}\quad
    \subfloat[]{\includegraphics[width=\columnwidth]{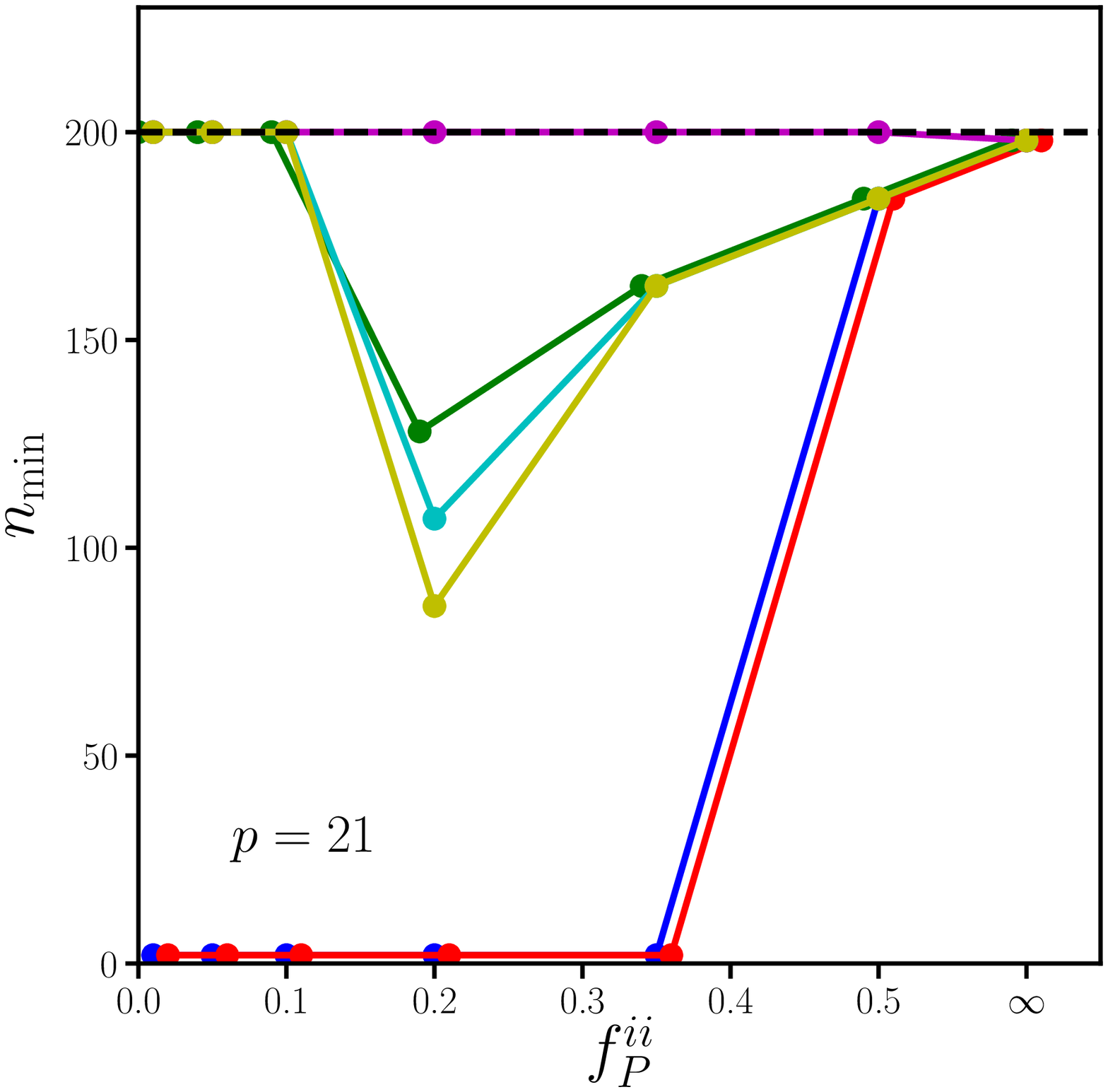}}
  \caption{\emph{Left panel}: The minimum number of simulations $n_{\mathrm{min}}$ allowed by the PTE, parameter-variance, and data-covariance-variance thresholds, as a function of the prior width $f_P^{ii}$ and adopting the more stringent thresholds (see text for details). We show the cases where the theory covariance is set to the true covariance (blue, second-lowest curve at $f_P^{ii}=0.4$), the true covariance scaled by 1.1 (red, lowest curve), the true covariance with its diagonal elements scaled by 1.1 (green, fourth-lowest curve at $f_P^{ii}=0.4$, overlapping with cyan), the true covariance with its off-diagonal elements scaled by 0.9 (cyan, second-highest curve at $f_P^{ii}=0.4$, overlapping with green) , the true covariance with its off-diagonal elements set to zero (magenta, highest curve), and the linear covariance (yellow, third-lowest curve at $f_P^{ii}=0.4$). The dashed horizontal line denotes the benchmark $n=200$ simulations required to achieve 10\% accuracy on the diagonal data covariance elements and 1\% loss in information on $A$ due to finite $n$ when the Jeffreys' prior is adopted. Note that some of the points have been horizontally offset for clarity, and there is a $\Delta n_{\mathrm{min}} = 7$ resolution when determining $n_{\mathrm{min}}$. \emph{Right panel}: Same as left panel, but adopting the less stringent PTE and parameter variance thresholds (see text). The ordering of the curves is now blue (second-lowest curve at $f_P^{ii}=0.4$), red (lowest curve), green (second-highest curve at $f_P^{ii}=0.2$), cyan (third-highest curve at $f_P^{ii}=0.2$), magenta (highest curve), and yellow (third-highest curve at $f_P^{ii}=0.2$). Note that there is no scale on the horizontal axis above $f_P^{ii} = 0.5$. Generally speaking, points on the left-hand side of these panels are constrained by giving reasonable model fits when the sampling distribution is the same as the null hypothesis, while points on the right-hand side are additionally constrained by having a reasonably precise data covariance matrix estimate.}
  \label{fig:minadmn}
\end{figure*}

Specifically, we choose two sets of thresholds; a `more stringent' set and a `less stringent' set. The more stringent set demands that the average PTE should not deviate from the true-null-hypothesis value 0.5 by more than $\pm 0.1$ when the model template $\boldsymbol{\mu}_0$ is correct (i.e. between the dot-dashed lines on Figure~\ref{fig:PTEplot}), that the posterior variance on the amplitude $A$ is not more than 10\% larger than the Jeffreys' prior value (i.e. below the dot-dashed line on Figure~\ref{fig:dvaratplot}), and that the standard deviation on the diagonal elements of the data covariance matrix in units of the true covariance is not greater than 10\% (i.e. below the dot-dashed line in Figure~\ref{fig:covcy}). The less stringent set relaxes these by allowing the average PTE to be $(0.5 \pm 0.3)$ i.e. between the dotted lines in Figure~\ref{fig:PTEplot}, while still below the dot-dashed lines in Figure~\ref{fig:dvaratplot} and Figure~\ref{fig:covcy}. We point out that these thresholds are arbitrary and one is free to specify any set of thresholds here.

We saw in Figure~\ref{fig:dvaratplot} that all of our models have a mean parameter variance lower than both the more stringent and less stringent thresholds, so for our particular choices of threshold this requirement plays no role in setting the value of $n_{\mathrm{min}}$. Instead, it is the PTE test and the covariance-variance test which determine how small we can make $n$. The smaller we make $n$, the more weight is put onto the theory covariance (with the transition controlled by the value of $f_P^{ii}$), and so the worse the PTE becomes since this covariance is not equal to the true value. If we counter this by increasing $f_P^{ii}$ to lessen the influence of the prior we can make the data covariance too noisy and $f_y^{ii}$ is too large. In the Jeffreys' prior we have seen that the benchmark number of simulations is 200 (see the discussion in Section~\ref{subsec:partition}), so if a combination of $n$ and $f_P^{ii}$ results in $n_{\mathrm{min}} \geq 200$ we conclude that the theory model $\mathbfss{C}_T$ is not good enough and we enforce $n_{\mathrm{min}} = 200$, independent of $p$.

In Figure~\ref{fig:minadmn} we plot $n_{\mathrm{min}}$ for each model choice, found by searching through all the values of $f_P^{ii}$ and $n$ we considered, with a resolution of $\Delta n = 7$. In the left panel we show the results for the more stringent thresholds, and in the right panel those for the less stringent thresholds. Note that since the averages all come with error bars there is some uncertainty in the determining the exact point at which the thresholds are satisfied for each $f_P^{ii}$ and $n$, but the values of $n_{\mathrm{min}}$ we obtained are stable to increasing the total number of simulations by 50\%.

When the more stringent thresholds are applied, the left panel of Figure~\ref{fig:minadmn} shows that only the covariance models closest to the truth can significantly reduce the number of simulations required. When the true covariance matrix is used (blue points), the number of simulations can be reduced to practically zero since the average PTE is always within the thresholds by construction. This is only untrue when $f_P^{ii}$ is large ($\gtrsim 0.5$), since then the noise on the data covariance exceeds the threshold for all values of $n \lesssim 200$ (as seen from the green curve in Figure~\ref{fig:covcy}). Similarly, the model with the true covariance scaled by 1.1 (red points) always has acceptable PTEs (see the top-middle panel of Figure~\ref{fig:PTEplot}), and so $n_{\mathrm{min}}$ can be very low as long as the data covariance is not too noisy (recall that Figure~\ref{fig:covcy} is independent of the model covariance choice).

If instead we choose a model with either the diagonal elements increased by 10\% or the off-diagonal elements decreased by 10\%, we can only obtain a modest decrease in $n_{\mathrm{min}}$, reducing the required number of simulations by roughly 10\% when a broad prior of $f_P^{ii} \approx 0.5$ is used. If a tighter prior is adopted the PTE becomes too large (see Figure~\ref{fig:PTEplot}), whereas if a weaker prior is used the variance in the data covariance becomes too large (Figure~\ref{fig:covcy}). If we throw away the off-diagonal elements completely the PTE becomes very poor even for the most uninformative IW prior, and so we can never improve $n_{\mathrm{min}}$ over the Jeffreys' result. The linear model has reasonable PTEs for broad priors with $f_P^{ii} \gtrsim 0.35$, allowing for a modest 20\% reduction in $n_{\mathrm{min}}$.

Thus, with these set of thresholds it appears that we can only obtain a significant reduction in the number of simulations that need to be run if the theory covariance is reasonably close to the truth in both its diagonal and off-diagonal elements. Even a small misestimation of the correlation matrix by 10\% gives poor $\chi^2$ values, allowing for only a 10\% reduction in $n_{\mathrm{min}}$.

If instead we relax the PTE threshold and allow the average PTE to be $0.5 \pm 0.3$, the models having 10\% underestimation of the correlation matrix become more acceptable and $n_{\mathrm{min}}$ comes down. There is a sweet spot of roughly $f_P^{ii} = 0.2 \pm 0.1$ (i.e. $20\% \pm 10\%$ confidence in the diagonal elements of the prior) where we can get down to $n_{\mathrm{min}} \approx 100 - 130$, almost a factor of two improvement. This value of $f_P^{ii}$ corresponds to $m-p-1 = 802$, i.e. the theoretical model is receiving roughly a factor of 7 more weight than the simulations in the data covariance matrix at this $n_{\mathrm{min}}$. A degradation of the PTE by 0.2 seems like a reasonable price to pay for this improvement. Further improvements are possible with the linear model due to the fortuitously good $\chi^2$ values that result from the competing influences of underestimated diagonal and off-diagonal elements (see the discussion in Section~\ref{subsec:PTEtests}), although the covariance-variance test is very approximate for this model, see Section~\ref{subsec:varCytests}.

To conclude this section, we have seen that substantial reductions in the number of simulations required for covariance matrix estimation are possible with a weakly informative prior centred on a model which gets the correlation matrix roughly correct. Tight priors are only helpful if the model covariance is close to the truth.

\subsection{Sensitivity to dimensionality}
\label{subsec:dim}

So far all our results have been obtained with a data vector of length $p=21$. This choice is fairly arbitrary, and so in this section we investigate the sensitivity of our results to changing $p$. The dimensionality can potentially be orders of magnitudes larger than this for planned surveys, but forecasting for this is challenging with our approach since we have to invert data covariance matrix tens of thousands of times to obtain the sample means used in our analysis. Instead, we re-ran our tests with $p=11$ and $p=31$ to get a rough idea of how this quantity impacts our results. The $k$-ranges are the same as in the $p=21$ results, with only the bin widths changed.

\begin{figure}
  \includegraphics[width=\columnwidth]{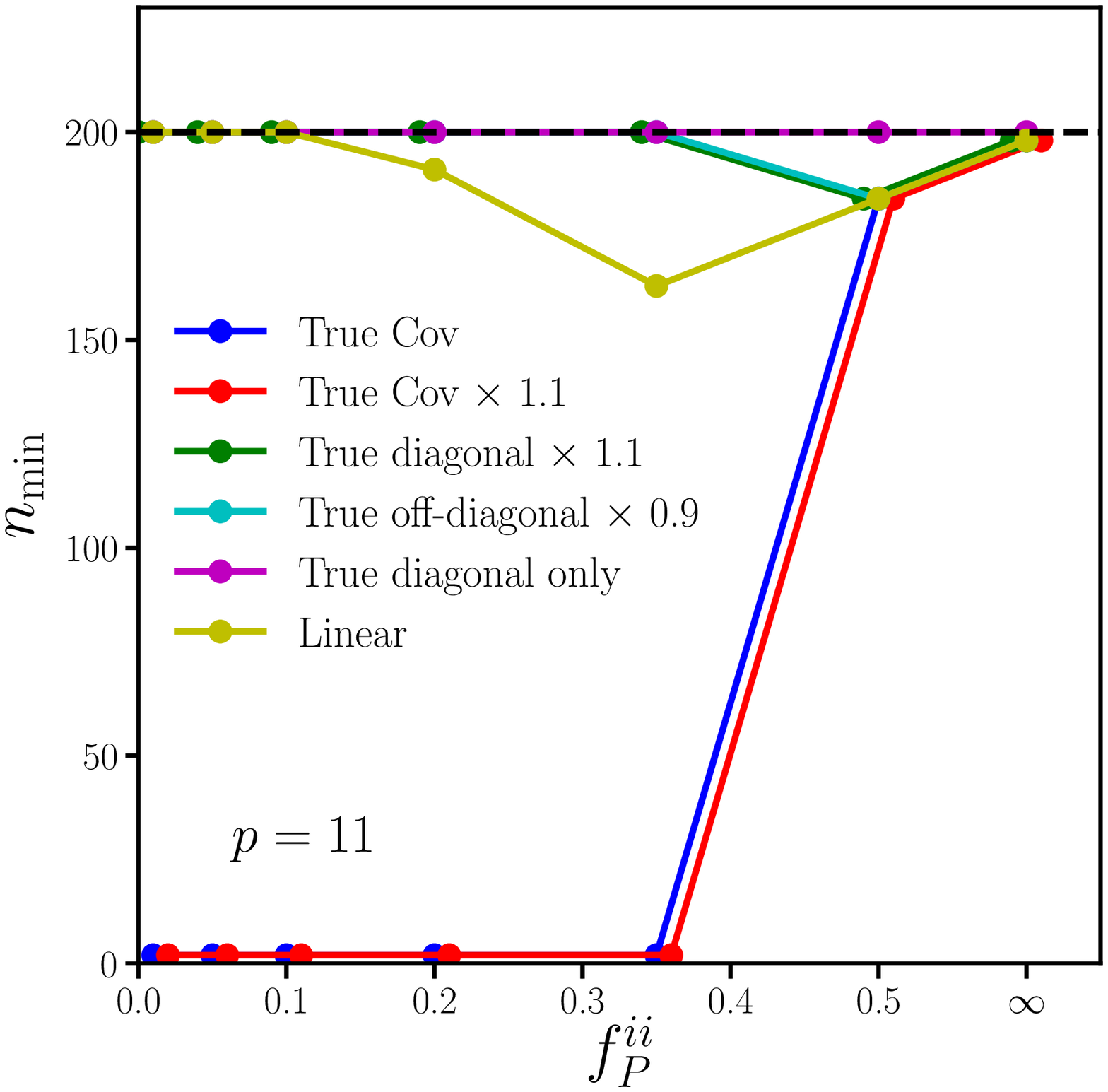}
  \includegraphics[width=\columnwidth]{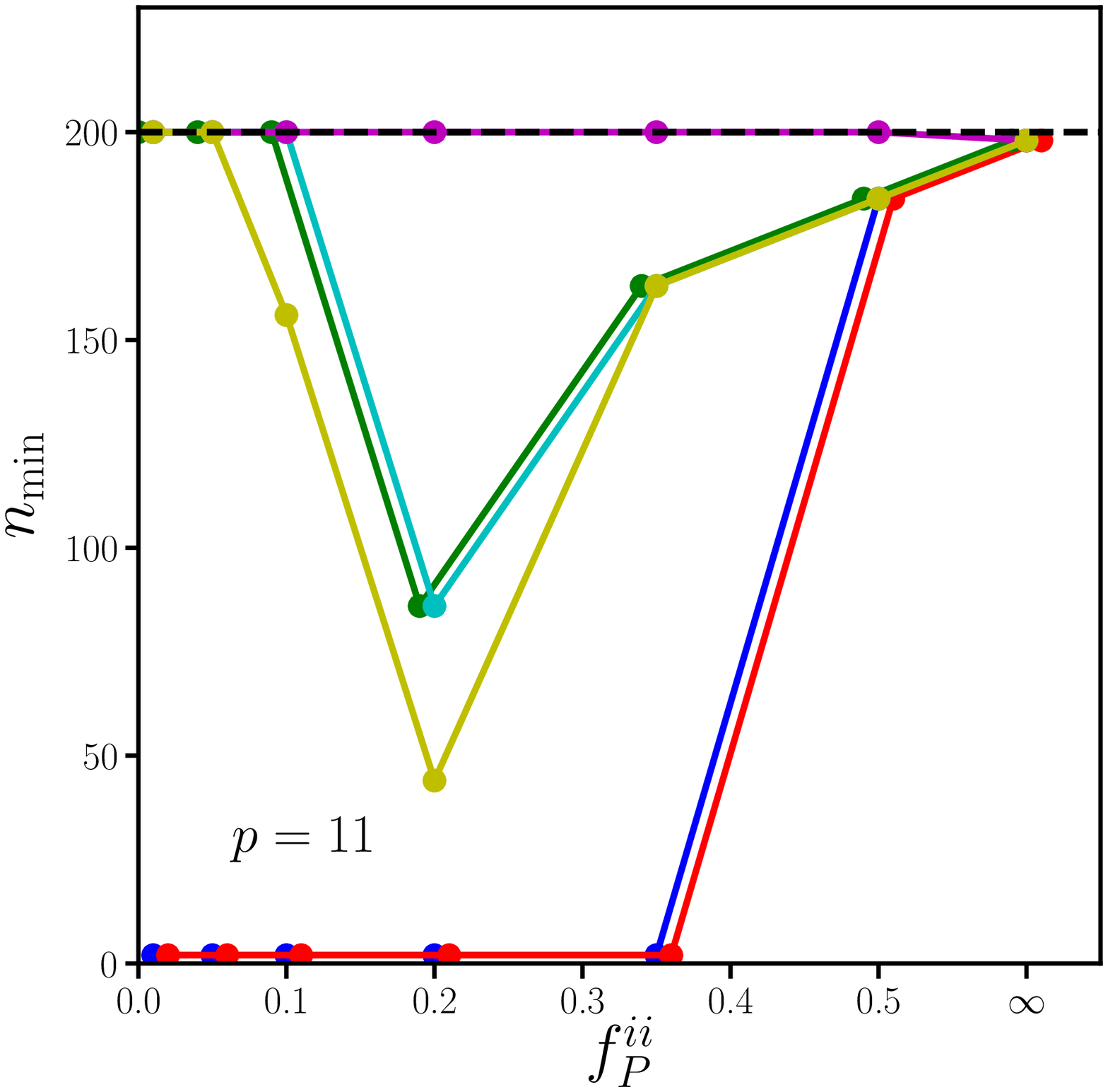}
  \caption{\emph{Top panel}: Same as left panel of Fig.~\ref{fig:minadmn} for $p=11$. \emph{Bottom panel}: Same as right panel of Fig.~\ref{fig:minadmn} for $p=11$.}
  \label{fig:p11minn}
\end{figure}

In Figure~\ref{fig:p11minn} we plot $n_{\mathrm{min}}$ for the various model covariance choices when $p=11$. The results with the more stringent thresholds (top panel) are very similar to the $p=21$ case, the only change being a small reduction in $n_{\mathrm{min}}$ for the linear model when $f_P^{ii} = 0.2$. More drastic changes occur when the less stringent thresholds are chosen (bottom panel), with the sweet-spot of $f_P^{ii} = 0.2$ now permitting over a factor two reduction in $n_{\mathrm{min}}$ for the 10\%-reduced-correlation models and the linear model. This is driven by a change to the PTE thresholds. We find that the average PTEs are closer to 0.5 for all the model choices, and can remain within the thresholds down to much lower $n$ than the $p=21$ case. This is because the width of the $\chi^2$ statistic when the null hypothesis is true is narrower when $p$ is larger since large fluctuations are less likely when there are a large number of statistically independent terms competing against each other in the sums in Equation~\eqref{eq:chi2}. This makes the PTE more sensitive to small changes in the error bars (and hence small changes in $\chi^2$) when the dimensionality is large~\citep{2017arXiv170801530D, 2018arXiv180410663T}. When $p$ is lowered this distribution broadens and we are less sensitive to poor estimation of the covariance matrix, allowing for lower $n_{\mathrm{min}}$. Using fewer $k$-bins also reduces the correlation between neighbouring bins, which typically makes the PTE less sensitive to misestimation of these correlations.

\begin{figure}
  \includegraphics[width=\columnwidth]{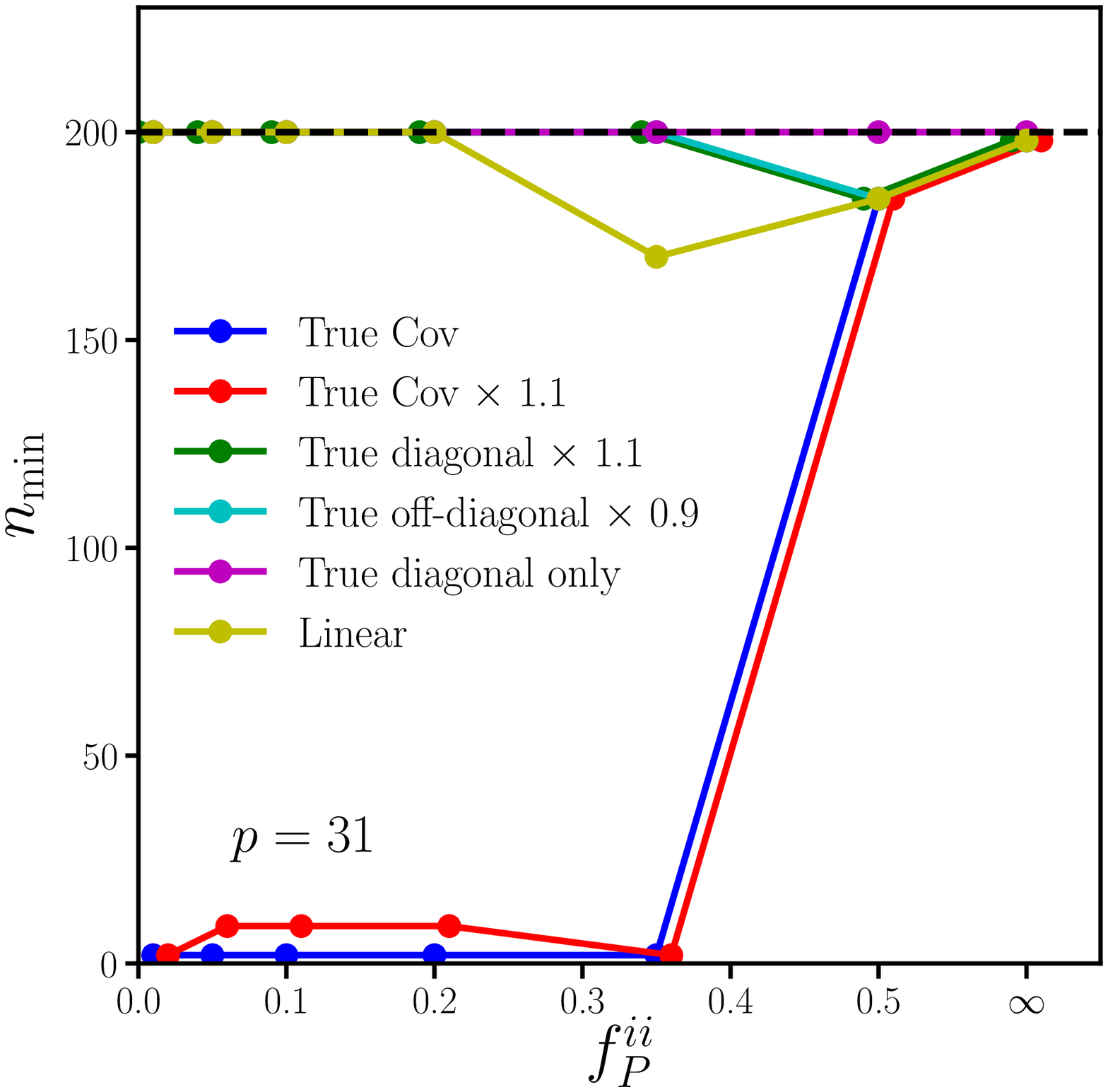}
  \includegraphics[width=\columnwidth]{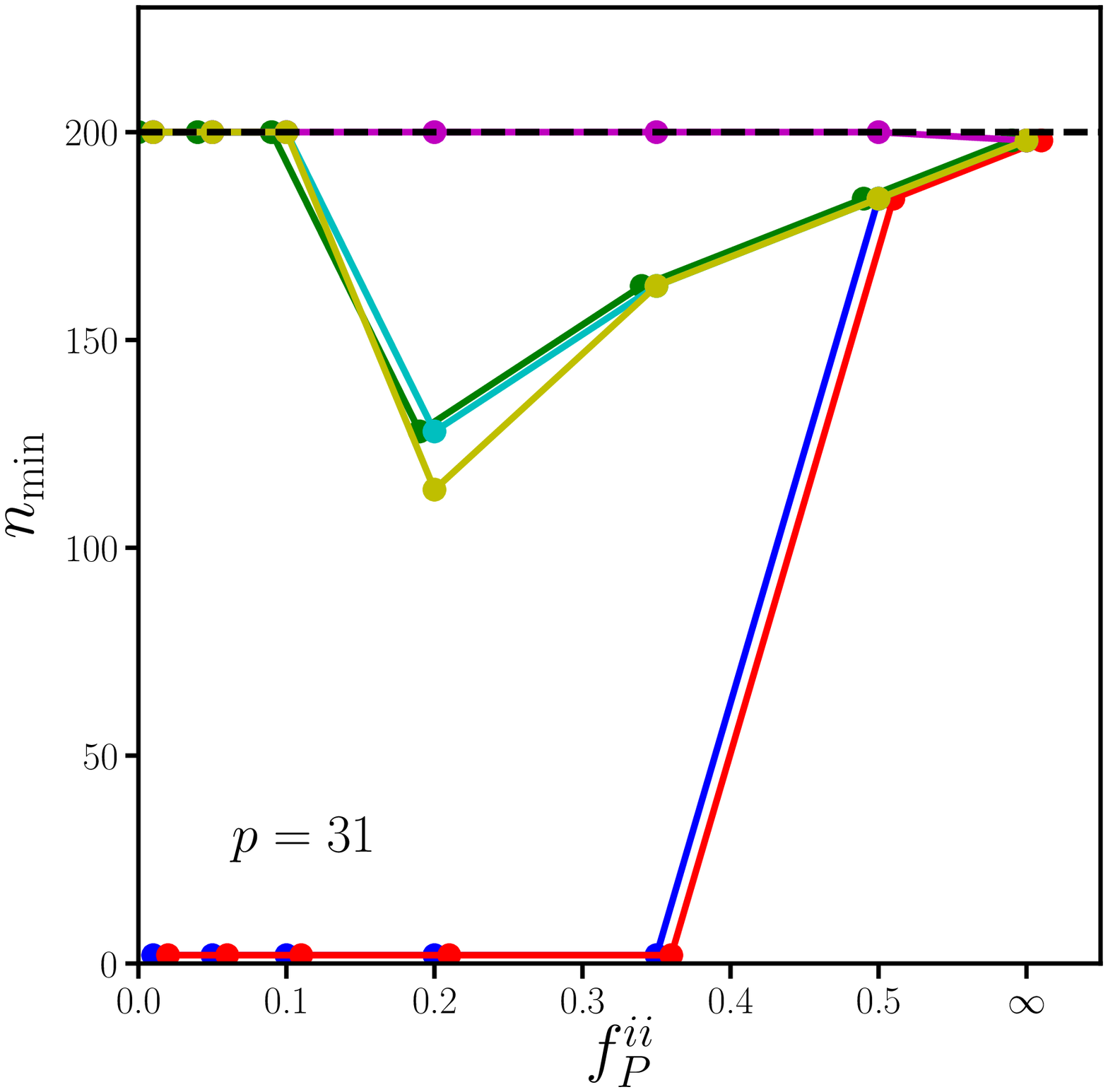}
  \caption{\emph{Top panel}: Same as left panel of Fig.~\ref{fig:minadmn} for $p=31$. \emph{Bottom panel}: Same as right panel of Fig.~\ref{fig:minadmn} for $p=31$.}
  \label{fig:p31minn}
\end{figure}

By the same token, when $p=31$ we are more sensitive to poor specification of the data covariance matrix and its correlation matrix, although the effect on $n_{\mathrm{min}}$ is slightly weaker, as we show in Figure~\ref{fig:p31minn}. When the thresholds are more stringent (top panel) there is little change from $p=21$ aside from a small degradation for the model where the true covariance is scaled by 1.1. When the thresholds are less stringent (bottom panel) there is a small increase in $n_{\mathrm{min}}$ at $f_P^{ii} = 0.2$ for the linear and scaled off-diagonal models compared to $p=21$. This is due to only a mild increase in the average PTE for this mis-specified models. We caution that these results are subject to noise in the average PTE.

The trend in $n_{\mathrm{min}}$ when we change $p$ is thus roughly as expected. We expect that there is also sensitivity to the range of $k$-scales we include, and indeed to how many redshift bins and summary statistics are folded in to the covariance matrix. For the very large values of $p$ which Stage-IV dark energy surveys are expected to produce, the PTE is likely to be quite sensitive to small changes in the data covariance matrix. This will make reducing $n_{\mathrm{min}}$ possible only for the weakest priors and most accurate models. The formalism we have presented in this work allows realistic covariance models to be assessed in a principled way to allow a sizeable reduction in $n_{\mathrm{min}}$.

For a survey with values of $p$ larger than those considered in this work, our recommendation is that the testing procedures outlined in this Section be followed, with a more realistic theory covariance matrix used to test the impact of high or low confidence in that theory on parameter errors and model fits. We have specialized to low values of $p$ to speed up some of the computations and to aid the interpretation of the results (for example, we have only considered real-space clustering in a single redshift bin), but the formalism presented here is completely general and could be used in a more realistic setting.

\section{Conclusions}
\label{sec:conc}

In this work we have introduced a new marginal likelihood for use in inferring cosmological parameters from surveys of cosmological large-scale structure. This likelihood is given in Equation~\eqref{eq:marglike}, the main results of this work, and has the form of a multivariate Student-$t$ distribution with a covariance matrix which linearly interpolates between a simulation-based estimate and a theoretical model.  This recovers the linear shrinkage model of~\citet{LEDOIT2004365} but in a Bayesian framework, and extends the approach of~\citet{2016MNRAS.456L.132S} by incorporating knowledge of what the true covariance should be, using an informative Inverse-Wishart prior. Our approach is motivated by the need to find new methods aimed at reducing the number of simulations $n_{\mathrm{min}}$ required for covariance estimation from the prohibitively large numbers forecast for planned dark energy experiments.

The marginal likelihood depends on a theoretical model and a degree-of-freedom parameter which determines the weight this model receives in the covariance matrix of the data. The weight can either be set from prior tests of the model on low-accuracy simulations or from internal predictions of the model's own accuracy. Alternatively it may be marginalised over in a Bayesian hierarchical model.

We have performed a thorough investigation of the impact on the inference process of different choices of the model, the weight, and the number of simulations forming the covariance matrix estimate. In Section~\ref{sec:results} we saw how having too much confidence in a poor model can result in a failure to pass a $\chi^2$ test even when the model for the mean of the data is correct. We saw that parameter variances are sensitive to these choices and how the noise in the hybrid estimator decreases when either strong confidence is placed on a model or the number of simulations is large. By placing thresholds on these quantities we were able to determine the minimum number of simulations required for our hybrid likelihood, finding that large reductions are possible if the model's correlation matrix is within roughly 10\% of the truth, and if one is willing to accept slightly poorer model fits.

Our approach provides a promising framework for combining theory and simulations in the likelihood, and resolves some of the confusion as to whether a `Hartlap correction' is required for hybrid covariance estimates~\citep{2007A&A...464..399H, 2013MNRAS.430.2200K}. Although the models we have considered have been simplistic, the methods we have employed should be useful in assessing whether any given model can be used to reduce $n_{\mathrm{min}}$.

One might wonder whether it is possible to include contributions to the covariance matrix not captured by simulations in our formalism, such as the super-sample covariance (SSC; \citealt{2013PhRvD..87l3504T}). In simplifying the marginalization over the covariance in Equation~\eqref{eq:likedef} we used that $ p( \mathbfss{C} | \boldsymbol{\mu},  \hat{\mathbfss{C}},  \mathbfss{C}_T) \propto  p( \hat{\mathbfss{C}} | \boldsymbol{\mu},  \mathbfss{C}) \, p(\mathbfss{C} | \boldsymbol{\mu} ,  \mathbfss{C}_T)$. We could imagine simply adding the SSC term to the simulation estimate $\hat{\mathbfss{C}}$, in which case the distribution $p( \hat{\mathbfss{C}} | \boldsymbol{\mu},  \mathbfss{C})$ becomes a shifted Wishart distribution. However, this means the marginalization can no longer be performed analytically. An alternative approach builds the SSC term into the model $\mathbfss{C}_T$ and treats the simulation estimate as biased. While this formally means that the distribution $p( \hat{\mathbfss{C}} | \boldsymbol{\mu},  \mathbfss{C})$ is no longer Wishart, we can always down-weight the influence of the simulations by increasing the theory weight, analogous to how we down-weighted our biased theoretical models. Since modern approaches of using theory-only covariance matrices are contained within our approach as a limit, it should be acceptable to include a simulation-based covariance estimate to furnish a theoretical model having an SSC term in order to improve the accuracy on non-linear scales. The weight placed on the simulation estimate would be quite small in scenarios where a theoretical covariance matrix was practically sufficient to meet the requirements of a survey~\citep{2018arXiv180704266B}.

Our approach is complementary to other methods aimed at combining theoretical and simulation-based covariance matrices~\citep{LEDOIT2004365, 2008MNRAS.389..766P, 2017MNRAS.466L..83J}, as well as methods which implement physically motivated approximations into simulations for covariance estimation~\citep{2013MNRAS.428.1036M, 2014MNRAS.439L..21K, 2015A&C....12..109H, 2016MNRAS.459.2327I, 2018arXiv180504537B}, and methods which compress the summary statistics into a reduced-dimensionality data vector~\citep{2017MNRAS.472.4244H}.

In conclusion, we have elucidated the role that prior confidence in models can play in reducing the computational demands placed on future dark energy experiments. With recent advances in understanding how estimated covariances impact likelihood functions and continuing progress in modelling the covariance, a hybrid approach embedded within a principled Bayesian framework such as that which we have described here will be a valuable tool for upcoming surveys.

\section*{Acknowledgements}

AH thanks Antony Lewis, Elena Sellentin and Joe Zuntz for useful conversations, and especially thanks Joachim Harnois-D\'{e}raps for providing the SLICS simulation results which were used in this work. The authors thank the referee for useful comments. AH is supported by an STFC Consolidated Grant. AT thanks the Royal Society for a Wolfson Research Merit Award, and the STFC for support from a Consolidated Grant.

%%%%%%%%%%%%%%%%%%%%%%%%%%%%%%%%%%%%%%%%%%%%%%%%%%

%%%%%%%%%%%%%%%%%%%% REFERENCES %%%%%%%%%%%%%%%%%%

% The best way to enter references is to use BibTeX:

\bibliographystyle{mnras}
\bibliography{references} % if your bibtex file is called example.bib

\begin{thebibliography}{}
\makeatletter
\relax
\def\mn@urlcharsother{\let\do\@makeother \do\$\do\&\do\#\do\^\do\_\do\%\do\~}
\def\mn@doi{\begingroup\mn@urlcharsother \@ifnextchar [ {\mn@doi@}
  {\mn@doi@[]}}
\def\mn@doi@[#1]#2{\def\@tempa{#1}\ifx\@tempa\@empty \href
  {http://dx.doi.org/#2} {doi:#2}\else \href {http://dx.doi.org/#2} {#1}\fi
  \endgroup}
\def\mn@eprint#1#2{\mn@eprint@#1:#2::\@nil}
\def\mn@eprint@arXiv#1{\href {http://arxiv.org/abs/#1} {{\tt arXiv:#1}}}
\def\mn@eprint@dblp#1{\href {http://dblp.uni-trier.de/rec/bibtex/#1.xml}
  {dblp:#1}}
\def\mn@eprint@#1:#2:#3:#4\@nil{\def\@tempa {#1}\def\@tempb {#2}\def\@tempc
  {#3}\ifx \@tempc \@empty \let \@tempc \@tempb \let \@tempb \@tempa \fi \ifx
  \@tempb \@empty \def\@tempb {arXiv}\fi \@ifundefined
  {mn@eprint@\@tempb}{\@tempb:\@tempc}{\expandafter \expandafter \csname
  mn@eprint@\@tempb\endcsname \expandafter{\@tempc}}}

\bibitem[\protect\citeauthoryear{Abbott et~al.,}{Abbott
  et~al.}{2018}]{2017arXiv170801530D}
Abbott T. M.~C.,  et~al., 2018, \mn@doi [Phys. Rev. D]
  {10.1103/PhysRevD.98.043526}, 98, 043526

\bibitem[\protect\citeauthoryear{{Alvarez}, {Niemi}  \& {Simpson}}{{Alvarez}
  et~al.}{2014}]{Alvarez}
{Alvarez} I.,  {Niemi} J.,   {Simpson} M.,  2014, in Song W.,  ed., Proceedings
  of 26th Annual Conference on Applied Statistics in Agriculture. New Prairie
  Press, pp 71--82

\bibitem[\protect\citeauthoryear{{Anderson}}{{Anderson}}{1958}]{Anderson}
{Anderson} T.~W.,  1958, {An Introduction to Multivariate Statistical
  Analysis}.
John Wiley \& Sons, Inc., Hoboken, NJ, USA

\bibitem[\protect\citeauthoryear{{Barreira} \& {Schmidt}}{{Barreira} \&
  {Schmidt}}{2017}]{2017JCAP...11..051B}
{Barreira} A.,  {Schmidt} F.,  2017, \mn@doi [\jcap]
  {10.1088/1475-7516/2017/11/051}, \href
  {http://adsabs.harvard.edu/abs/2017JCAP...11..051B} {11, 051}

\bibitem[\protect\citeauthoryear{{Barreira}, {Krause}  \& {Schmidt}}{{Barreira}
  et~al.}{2018}]{2018arXiv180704266B}
{Barreira} A.,  {Krause} E.,   {Schmidt} F.,  2018, preprint, \href
  {http://adsabs.harvard.edu/abs/2018arXiv180704266B} {} (\mn@eprint {arXiv}
  {1807.04266})

\bibitem[\protect\citeauthoryear{{Berger} \& {Stein}}{{Berger} \&
  {Stein}}{2019}]{2018arXiv180504537B}
{Berger} P.,  {Stein} G.,  2019, \mn@doi [\mnras] {10.1093/mnras/sty2949},
  \href {http://adsabs.harvard.edu/abs/2019MNRAS.482.2861B} {482, 2861}

\bibitem[\protect\citeauthoryear{{Bernardeau}, {Colombi}, {Gazta{\~n}aga}  \&
  {Scoccimarro}}{{Bernardeau} et~al.}{2002}]{2002PhR...367....1B}
{Bernardeau} F.,  {Colombi} S.,  {Gazta{\~n}aga} E.,   {Scoccimarro} R.,  2002,
  \mn@doi [\physrep] {10.1016/S0370-1573(02)00135-7}, \href
  {http://adsabs.harvard.edu/abs/2002PhR...367....1B} {367, 1}

\bibitem[\protect\citeauthoryear{{Bertolini}, {Schutz}, {Solon}, {Walsh}  \&
  {Zurek}}{{Bertolini} et~al.}{2016}]{2016PhRvD..93l3505B}
{Bertolini} D.,  {Schutz} K.,  {Solon} M.~P.,  {Walsh} J.~R.,   {Zurek} K.~M.,
  2016, \mn@doi [\prd] {10.1103/PhysRevD.93.123505}, \href
  {http://adsabs.harvard.edu/abs/2016PhRvD..93l3505B} {93, 123505}

\bibitem[\protect\citeauthoryear{{Blot}, {Corasaniti}, {Alimi}, {Reverdy}  \&
  {Rasera}}{{Blot} et~al.}{2015}]{2015MNRAS.446.1756B}
{Blot} L.,  {Corasaniti} P.~S.,  {Alimi} J.-M.,  {Reverdy} V.,   {Rasera} Y.,
  2015, \mn@doi [\mnras] {10.1093/mnras/stu2190}, \href
  {http://adsabs.harvard.edu/abs/2015MNRAS.446.1756B} {446, 1756}

\bibitem[\protect\citeauthoryear{{Cooray} \& {Sheth}}{{Cooray} \&
  {Sheth}}{2002}]{2002PhR...372....1C}
{Cooray} A.,  {Sheth} R.,  2002, \mn@doi [\physrep]
  {10.1016/S0370-1573(02)00276-4}, \href
  {http://adsabs.harvard.edu/abs/2002PhR...372....1C} {372, 1}

\bibitem[\protect\citeauthoryear{{Cooray}, {Hu}  \&
  {Miralda-Escud{\'e}}}{{Cooray} et~al.}{2000}]{2000ApJ...535L...9C}
{Cooray} A.,  {Hu} W.,   {Miralda-Escud{\'e}} J.,  2000, \mn@doi [\apjl]
  {10.1086/312696}, \href {http://adsabs.harvard.edu/abs/2000ApJ...535L...9C}
  {535, L9}

\bibitem[\protect\citeauthoryear{{Dodelson} \& {Schneider}}{{Dodelson} \&
  {Schneider}}{2013}]{2013PhRvD..88f3537D}
{Dodelson} S.,  {Schneider} M.~D.,  2013, \mn@doi [\prd]
  {10.1103/PhysRevD.88.063537}, \href
  {http://adsabs.harvard.edu/abs/2013PhRvD..88f3537D} {88, 063537}

\bibitem[\protect\citeauthoryear{{Friedrich} \& {Eifler}}{{Friedrich} \&
  {Eifler}}{2018}]{2018MNRAS.473.4150F}
{Friedrich} O.,  {Eifler} T.,  2018, \mn@doi [\mnras] {10.1093/mnras/stx2566},
  \href {http://adsabs.harvard.edu/abs/2018MNRAS.473.4150F} {473, 4150}

\bibitem[\protect\citeauthoryear{{Gelman}, {Carlin}, {Stern}, {Dunson},
  {Vehtari}  \& {Rubin}}{{Gelman} et~al.}{1995}]{Gelman}
{Gelman} A.,  {Carlin} J.,  {Stern} H.,  {Dunson} D.,  {Vehtari} A.,   {Rubin}
  D.,  1995, {Bayesian Data Analysis}.
Chapman and Hall, CRC Press, Boca Raton, FL, USA

\bibitem[\protect\citeauthoryear{{Gupta} \& {Nagar}}{{Gupta} \&
  {Nagar}}{2000}]{Gupta}
{Gupta} A.~K.,  {Nagar} D.~K.,  2000, {Matrix Variate Distributions}.
Chapman and Hall, CRC Press, Boca Raton, FL, USA

\bibitem[\protect\citeauthoryear{{Hahn}, {Beutler}, {Sinha}, {Berlind}, {Ho}
  \& {Hogg}}{{Hahn} et~al.}{2018}]{2018arXiv180306348H}
{Hahn} C.,  {Beutler} F.,  {Sinha} M.,  {Berlind} A.,  {Ho} S.,   {Hogg} D.~W.,
   2018, preprint, \href {http://adsabs.harvard.edu/abs/2018arXiv180306348H} {}
  (\mn@eprint {arXiv} {1803.06348})

\bibitem[\protect\citeauthoryear{{Hamimeche} \& {Lewis}}{{Hamimeche} \&
  {Lewis}}{2008}]{2008PhRvD..77j3013H}
{Hamimeche} S.,  {Lewis} A.,  2008, \mn@doi [\prd]
  {10.1103/PhysRevD.77.103013}, \href
  {http://adsabs.harvard.edu/abs/2008PhRvD..77j3013H} {77, 103013}

\bibitem[\protect\citeauthoryear{{Hamimeche} \& {Lewis}}{{Hamimeche} \&
  {Lewis}}{2009}]{2009PhRvD..79h3012H}
{Hamimeche} S.,  {Lewis} A.,  2009, \mn@doi [\prd]
  {10.1103/PhysRevD.79.083012}, \href
  {http://adsabs.harvard.edu/abs/2009PhRvD..79h3012H} {79, 083012}

\bibitem[\protect\citeauthoryear{{Harnois-D{\'e}raps} \& {van
  Waerbeke}}{{Harnois-D{\'e}raps} \& {van
  Waerbeke}}{2015}]{2015MNRAS.450.2857H}
{Harnois-D{\'e}raps} J.,  {van Waerbeke} L.,  2015, \mn@doi [\mnras]
  {10.1093/mnras/stv794}, \href
  {http://adsabs.harvard.edu/abs/2015MNRAS.450.2857H} {450, 2857}

\bibitem[\protect\citeauthoryear{{Harnois-D{\'e}raps}, {Vafaei}  \& {Van
  Waerbeke}}{{Harnois-D{\'e}raps} et~al.}{2012}]{2012MNRAS.426.1262H}
{Harnois-D{\'e}raps} J.,  {Vafaei} S.,   {Van Waerbeke} L.,  2012, \mn@doi
  [\mnras] {10.1111/j.1365-2966.2012.21624.x}, \href
  {http://adsabs.harvard.edu/abs/2012MNRAS.426.1262H} {426, 1262}

\bibitem[\protect\citeauthoryear{{Harnois-D{\'e}raps}
  et~al.,}{{Harnois-D{\'e}raps} et~al.}{2018}]{2018arXiv180504511H}
{Harnois-D{\'e}raps} J.,  et~al., 2018, \mn@doi [\mnras]
  {10.1093/mnras/sty2319}, \href
  {http://adsabs.harvard.edu/abs/2018MNRAS.481.1337H} {481, 1337}

\bibitem[\protect\citeauthoryear{{Hartlap}, {Simon}  \& {Schneider}}{{Hartlap}
  et~al.}{2007}]{2007A&A...464..399H}
{Hartlap} J.,  {Simon} P.,   {Schneider} P.,  2007, \mn@doi [\aap]
  {10.1051/0004-6361:20066170}, \href
  {http://adsabs.harvard.edu/abs/2007A%26A...464..399H} {464, 399}

\bibitem[\protect\citeauthoryear{{Hartlap}, {Schrabback}, {Simon}  \&
  {Schneider}}{{Hartlap} et~al.}{2009}]{2009A&A...504..689H}
{Hartlap} J.,  {Schrabback} T.,  {Simon} P.,   {Schneider} P.,  2009, \mn@doi
  [\aap] {10.1051/0004-6361/200911697}, \href
  {http://adsabs.harvard.edu/abs/2009A%26A...504..689H} {504, 689}

\bibitem[\protect\citeauthoryear{{Heavens}, {Sellentin}, {de Mijolla}  \&
  {Vianello}}{{Heavens} et~al.}{2017}]{2017MNRAS.472.4244H}
{Heavens} A.~F.,  {Sellentin} E.,  {de Mijolla} D.,   {Vianello} A.,  2017,
  \mn@doi [\mnras] {10.1093/mnras/stx2326}, \href
  {http://adsabs.harvard.edu/abs/2017MNRAS.472.4244H} {472, 4244}

\bibitem[\protect\citeauthoryear{{Heitmann}, {Lawrence}, {Kwan}, {Habib}  \&
  {Higdon}}{{Heitmann} et~al.}{2014}]{2014ApJ...780..111H}
{Heitmann} K.,  {Lawrence} E.,  {Kwan} J.,  {Habib} S.,   {Higdon} D.,  2014,
  \mn@doi [\apj] {10.1088/0004-637X/780/1/111}, \href
  {http://adsabs.harvard.edu/abs/2014ApJ...780..111H} {780, 111}

\bibitem[\protect\citeauthoryear{{Hinshaw} et~al.,}{{Hinshaw}
  et~al.}{2013}]{2013ApJS..208...19H}
{Hinshaw} G.,  et~al., 2013, \mn@doi [\apjs] {10.1088/0067-0049/208/2/19},
  \href {http://adsabs.harvard.edu/abs/2013ApJS..208...19H} {208, 19}

\bibitem[\protect\citeauthoryear{{Howlett}, {Lewis}, {Hall}  \&
  {Challinor}}{{Howlett} et~al.}{2012}]{2012JCAP...04..027H}
{Howlett} C.,  {Lewis} A.,  {Hall} A.,   {Challinor} A.,  2012, \mn@doi [\jcap]
  {10.1088/1475-7516/2012/04/027}, \href
  {http://adsabs.harvard.edu/abs/2012JCAP...04..027H} {4, 027}

\bibitem[\protect\citeauthoryear{{Howlett}, {Manera}  \& {Percival}}{{Howlett}
  et~al.}{2015}]{2015A&C....12..109H}
{Howlett} C.,  {Manera} M.,   {Percival} W.~J.,  2015, \mn@doi [Astronomy and
  Computing] {10.1016/j.ascom.2015.07.003}, \href
  {http://adsabs.harvard.edu/abs/2015A%26C....12..109H} {12, 109}

\bibitem[\protect\citeauthoryear{{Izard}, {Crocce}  \& {Fosalba}}{{Izard}
  et~al.}{2016}]{2016MNRAS.459.2327I}
{Izard} A.,  {Crocce} M.,   {Fosalba} P.,  2016, \mn@doi [\mnras]
  {10.1093/mnras/stw797}, \href
  {http://adsabs.harvard.edu/abs/2016MNRAS.459.2327I} {459, 2327}

\bibitem[\protect\citeauthoryear{{Joachimi}}{{Joachimi}}{2017}]{2017MNRAS.466L..83J}
{Joachimi} B.,  2017, \mn@doi [\mnras] {10.1093/mnrasl/slw240}, \href
  {http://adsabs.harvard.edu/abs/2017MNRAS.466L..83J} {466, L83}

\bibitem[\protect\citeauthoryear{{Kilbinger} et~al.,}{{Kilbinger}
  et~al.}{2013}]{2013MNRAS.430.2200K}
{Kilbinger} M.,  et~al., 2013, \mn@doi [\mnras] {10.1093/mnras/stt041}, \href
  {http://adsabs.harvard.edu/abs/2013MNRAS.430.2200K} {430, 2200}

\bibitem[\protect\citeauthoryear{{Kitaura}, {Yepes}  \& {Prada}}{{Kitaura}
  et~al.}{2014}]{2014MNRAS.439L..21K}
{Kitaura} F.-S.,  {Yepes} G.,   {Prada} F.,  2014, \mn@doi [\mnras]
  {10.1093/mnrasl/slt172}, \href
  {http://adsabs.harvard.edu/abs/2014MNRAS.439L..21K} {439, L21}

\bibitem[\protect\citeauthoryear{{Leclercq}}{{Leclercq}}{2018}]{2018arXiv180507152L}
{Leclercq} F.,  2018, \mn@doi [\prd] {10.1103/PhysRevD.98.063511}, \href
  {http://adsabs.harvard.edu/abs/2018PhRvD..98f3511L} {98, 063511}

\bibitem[\protect\citeauthoryear{Ledoit \& Wolf}{Ledoit \&
  Wolf}{2004}]{LEDOIT2004365}
Ledoit O.,  Wolf M.,  2004, \mn@doi [Journal of Multivariate Analysis]
  {https://doi.org/10.1016/S0047-259X(03)00096-4}, 88, 365

\bibitem[\protect\citeauthoryear{{Lewis}, {Challinor}  \& {Lasenby}}{{Lewis}
  et~al.}{2000}]{2000ApJ...538..473L}
{Lewis} A.,  {Challinor} A.,   {Lasenby} A.,  2000, \mn@doi [\apj]
  {10.1086/309179}, \href {http://adsabs.harvard.edu/abs/2000ApJ...538..473L}
  {538, 473}

\bibitem[\protect\citeauthoryear{{Manera} et~al.,}{{Manera}
  et~al.}{2013}]{2013MNRAS.428.1036M}
{Manera} M.,  et~al., 2013, \mn@doi [\mnras] {10.1093/mnras/sts084}, \href
  {http://adsabs.harvard.edu/abs/2013MNRAS.428.1036M} {428, 1036}

\bibitem[\protect\citeauthoryear{{McCarthy}, {Schaye}, {Bird}  \& {Le
  Brun}}{{McCarthy} et~al.}{2017}]{2017MNRAS.465.2936M}
{McCarthy} I.~G.,  {Schaye} J.,  {Bird} S.,   {Le Brun} A.~M.~C.,  2017,
  \mn@doi [\mnras] {10.1093/mnras/stw2792}, \href
  {http://adsabs.harvard.edu/abs/2017MNRAS.465.2936M} {465, 2936}

\bibitem[\protect\citeauthoryear{{Mohammed}, {Seljak}  \& {Vlah}}{{Mohammed}
  et~al.}{2017}]{2017MNRAS.466..780M}
{Mohammed} I.,  {Seljak} U.,   {Vlah} Z.,  2017, \mn@doi [\mnras]
  {10.1093/mnras/stw3196}, \href
  {http://adsabs.harvard.edu/abs/2017MNRAS.466..780M} {466, 780}

\bibitem[\protect\citeauthoryear{{Peacock} \& {Smith}}{{Peacock} \&
  {Smith}}{2000}]{2000MNRAS.318.1144P}
{Peacock} J.~A.,  {Smith} R.~E.,  2000, \mn@doi [\mnras]
  {10.1046/j.1365-8711.2000.03779.x}, \href
  {http://adsabs.harvard.edu/abs/2000MNRAS.318.1144P} {318, 1144}

\bibitem[\protect\citeauthoryear{{Planck Collaboration} et~al.,}{{Planck
  Collaboration} et~al.}{2014}]{2014A&A...571A..15P}
{Planck Collaboration} et~al., 2014, \mn@doi [\aap]
  {10.1051/0004-6361/201321573}, \href
  {http://adsabs.harvard.edu/abs/2014A%26A...571A..15P} {571, A15}

\bibitem[\protect\citeauthoryear{{Pope} \& {Szapudi}}{{Pope} \&
  {Szapudi}}{2008}]{2008MNRAS.389..766P}
{Pope} A.~C.,  {Szapudi} I.,  2008, \mn@doi [\mnras]
  {10.1111/j.1365-2966.2008.13561.x}, \href
  {http://adsabs.harvard.edu/abs/2008MNRAS.389..766P} {389, 766}

\bibitem[\protect\citeauthoryear{Schafer \& Freeman}{Schafer \&
  Freeman}{2012}]{10.1007/978-1-4614-3520-4_1}
Schafer C.~M.,  Freeman P.~E.,  2012, in Feigelson E.~D.,  Babu G.~J.,  eds,
  Statistical Challenges in Modern Astronomy V. Springer New York, New York,
  NY, pp 3--19

\bibitem[\protect\citeauthoryear{{Scoccimarro}}{{Scoccimarro}}{2000}]{2000ApJ...544..597S}
{Scoccimarro} R.,  2000, \mn@doi [\apj] {10.1086/317248}, \href
  {http://adsabs.harvard.edu/abs/2000ApJ...544..597S} {544, 597}

\bibitem[\protect\citeauthoryear{{Seljak}}{{Seljak}}{2000}]{2000MNRAS.318..203S}
{Seljak} U.,  2000, \mn@doi [\mnras] {10.1046/j.1365-8711.2000.03715.x}, \href
  {http://adsabs.harvard.edu/abs/2000MNRAS.318..203S} {318, 203}

\bibitem[\protect\citeauthoryear{{Sellentin} \& {Heavens}}{{Sellentin} \&
  {Heavens}}{2016}]{2016MNRAS.456L.132S}
{Sellentin} E.,  {Heavens} A.~F.,  2016, \mn@doi [\mnras]
  {10.1093/mnrasl/slv190}, \href
  {http://adsabs.harvard.edu/abs/2016MNRAS.456L.132S} {456, L132}

\bibitem[\protect\citeauthoryear{{Sellentin} \& {Heavens}}{{Sellentin} \&
  {Heavens}}{2017}]{2017MNRAS.464.4658S}
{Sellentin} E.,  {Heavens} A.~F.,  2017, \mn@doi [\mnras]
  {10.1093/mnras/stw2697}, \href
  {http://adsabs.harvard.edu/abs/2017MNRAS.464.4658S} {464, 4658}

\bibitem[\protect\citeauthoryear{{Sellentin} \& {Heavens}}{{Sellentin} \&
  {Heavens}}{2018}]{2018MNRAS.473.2355S}
{Sellentin} E.,  {Heavens} A.~F.,  2018, \mn@doi [\mnras]
  {10.1093/mnras/stx2491}, \href
  {http://adsabs.harvard.edu/abs/2018MNRAS.473.2355S} {473, 2355}

\bibitem[\protect\citeauthoryear{{Simpson}, {Harnois-D{\'e}raps}, {Heymans},
  {Jimenez}, {Joachimi}  \& {Verde}}{{Simpson}
  et~al.}{2016}]{2016MNRAS.456..278S}
{Simpson} F.,  {Harnois-D{\'e}raps} J.,  {Heymans} C.,  {Jimenez} R.,
  {Joachimi} B.,   {Verde} L.,  2016, \mn@doi [\mnras] {10.1093/mnras/stv2474},
  \href {http://adsabs.harvard.edu/abs/2016MNRAS.456..278S} {456, 278}

\bibitem[\protect\citeauthoryear{{Takada} \& {Hu}}{{Takada} \&
  {Hu}}{2013}]{2013PhRvD..87l3504T}
{Takada} M.,  {Hu} W.,  2013, \mn@doi [\prd] {10.1103/PhysRevD.87.123504},
  \href {http://adsabs.harvard.edu/abs/2013PhRvD..87l3504T} {87, 123504}

\bibitem[\protect\citeauthoryear{{Takahashi} et~al.,}{{Takahashi}
  et~al.}{2009}]{2009ApJ...700..479T}
{Takahashi} R.,  et~al., 2009, \mn@doi [\apj] {10.1088/0004-637X/700/1/479},
  \href {http://adsabs.harvard.edu/abs/2009ApJ...700..479T} {700, 479}

\bibitem[\protect\citeauthoryear{{Takahashi} et~al.,}{{Takahashi}
  et~al.}{2011}]{2011ApJ...726....7T}
{Takahashi} R.,  et~al., 2011, \mn@doi [\apj] {10.1088/0004-637X/726/1/7},
  \href {http://adsabs.harvard.edu/abs/2011ApJ...726....7T} {726, 7}

\bibitem[\protect\citeauthoryear{{Taylor}, {Joachimi}  \& {Kitching}}{{Taylor}
  et~al.}{2013}]{2013MNRAS.432.1928T}
{Taylor} A.,  {Joachimi} B.,   {Kitching} T.,  2013, \mn@doi [\mnras]
  {10.1093/mnras/stt270}, \href
  {http://adsabs.harvard.edu/abs/2013MNRAS.432.1928T} {432, 1928}

\bibitem[\protect\citeauthoryear{{Troxel} et~al.,}{{Troxel}
  et~al.}{2018}]{2018arXiv180410663T}
{Troxel} M.~A.,  et~al., 2018, \mn@doi [\mnras] {10.1093/mnras/sty1889}, \href
  {http://adsabs.harvard.edu/abs/2018MNRAS.479.4998T} {479, 4998}

\bibitem[\protect\citeauthoryear{{Weyant}, {Schafer}  \& {Wood-Vasey}}{{Weyant}
  et~al.}{2013}]{2013ApJ...764..116W}
{Weyant} A.,  {Schafer} C.,   {Wood-Vasey} W.~M.,  2013, \mn@doi [\apj]
  {10.1088/0004-637X/764/2/116}, \href
  {http://adsabs.harvard.edu/abs/2013ApJ...764..116W} {764, 116}

\bibitem[\protect\citeauthoryear{{White} \& {Padmanabhan}}{{White} \&
  {Padmanabhan}}{2015}]{2015JCAP...12..058W}
{White} M.,  {Padmanabhan} N.,  2015, \mn@doi [\jcap]
  {10.1088/1475-7516/2015/12/058}, \href
  {http://adsabs.harvard.edu/abs/2015JCAP...12..058W} {12, 058}

\bibitem[\protect\citeauthoryear{{Wittman}, {Bhaskar}  \& {Tobin}}{{Wittman}
  et~al.}{2016}]{2016MNRAS.457.4005W}
{Wittman} D.,  {Bhaskar} R.,   {Tobin} R.,  2016, \mn@doi [\mnras]
  {10.1093/mnras/stw261}, \href
  {http://adsabs.harvard.edu/abs/2016MNRAS.457.4005W} {457, 4005}

\makeatother
\end{thebibliography}

% Alternatively you could enter them by hand, like this:
% This method is tedious and prone to error if you have lots of references
%\begin{thebibliography}{99}
%\bibitem[\protect\citeauthoryear{Author}{2012}]{Author2012}
%Author A.~N., 2013, Journal of Improbable Astronomy, 1, 1
%\bibitem[\protect\citeauthoryear{Others}{2013}]{Others2013}
%Others S., 2012, Journal of Interesting Stuff, 17, 198
%\end{thebibliography}

%%%%%%%%%%%%%%%%%%%%%%%%%%%%%%%%%%%%%%%%%%%%%%%%%%

%%%%%%%%%%%%%%%%% APPENDICES %%%%%%%%%%%%%%%%%%%%%

\appendix

\section{Inverse-Wishart mixture prior}
\label{app:IWmix}

The IW prior allows the marginalization over the unknown covariance matrix to be performed analytically. The price we pay for this however is a loss in flexibility, as once we fix the mean of the prior to equal the theoretical model there is only one remaining free parameter, the degree-of-freedom $m$. We can remedy this by using a mixture of $N$ Inverse Wishart distributions, each having a different scale matrix and degree-of-freedom parameter, with density given by
\begin{equation}
  p(\mathbfss{C} | \mathbfss{C}_T) = \sum_{k=1}^N w_k p_k(\mathbfss{C} | \bar{\mathbfss{C}}_k(m_k - p - 1), m_k),
  \label{eq:IWmix}
\end{equation}
where each $p_k$ is an IW distribution with mean $\bar{\mathbfss{C}}_k$ and degree-of-freedom $m_k$, and the weights must satisfy
\begin{equation}
  \sum_{k=1}^N w_k = 1
\end{equation}
to ensure the prior integrates to unity. The mean of the prior is
\begin{equation}
  \bar{\mathbfss{C}} = \sum_{k=1}^N w_k \bar{\mathbfss{C}}_k,
  \label{eq:mixmean}
\end{equation}
which we can set equal to the theoretical model $\mathbfss{C}_T$, as in the single IW case.

With the IW mixture prior we can again marginalize over $\mathbfss{C}$ analytically and derive the marginal likelihood, which is now a weighted sum of multivariate Student-$t$ distributions, each with a common mean $\boldsymbol{\mu}$. The data covariance of this marginal likelihood is
\begin{equation}
  \mathbfss{C}_{\mathbfit{y}} = \sum_{k=1}^N w_k' \mathbfss{C}_{\mathbfit{y},k},
  \label{eq:Cymix}
\end{equation}
where $w_k'$ are new weights related to the $w_k$ via elementary functions (and required to sum to unity), while $\mathbfss{C}_{\mathbfit{y},k}$ are the covariances of the individual multivariate-$t$ distributions in the mixture.

Similarly, we can derive the posterior distribution of an amplitude parameter exactly, which takes the form of a weighted sum of Student-$t$ distributions, each with a mean given by $\mu_k = \boldsymbol{\mu}_0^\intercal \mathbfss{C}_{\mathbfit{y},k}^{-1} \mathbfit{y}/\boldsymbol{\mu}_0^\intercal \mathbfss{C}_{\mathbfit{y},k}^{-1}\boldsymbol{\mu}_0$. If we define the weights of this mixture by $w_k''$, related to the original prior weights by elementary functions, the posterior mean and variance of $A$ are now
\begin{align}
  &\langle A \rangle = \sum_{k=1}^N w_k'' \mu_k \nonumber \\
  &\mathrm{var}(A) = \sum_{k=1}^N w_k'' \mathrm{var}_k(A) + \sum_{k=1}^N w_k'' \left(\mu_k - \sum_{m=1}^N w_m'' \mu_m \right)^2,
  \label{eq:Astatsmix}
\end{align}
where $\mathrm{var}_k(A)$ is the variance of each individual component, given by Equation~\eqref{eq:Astats} with the replacements $m \rightarrow m_k$ and $\mathbfss{C}_{\mathbfit{y}} \rightarrow \mathbfss{C}_{\mathbfit{y},k}$. The parameter variance is thus a sum of the weighted individual variances of the mixture components and an `intrinsic' variance coming from the scatter in peak locations within the mixture. It is also easy to show that the sampling distribution of the $\chi^2$ test statistic defined in Equation~\eqref{eq:chi2} with $\mathbfss{C}_T = \mathbfss{C}_0$ is independent of the unknown model parameters, following the derivation in Appendix~\ref{app:chi2dist}.

It thus remains for us to specify the parameters of each mixture component $\bar{\mathbfss{C}}_k$ and $m_k$, and the weights $w_k$. We would like to control the variance of different blocks of the covariance matrix. These blocks could be blocks in scale or redshift or particular combinations of the summary statistics (e.g. the covariance of the position-shear correlation function with the shear-shear correlation function), and the weighted sum of the blocks must equal the theoretical model, i.e. $\bar{\mathbfss{C}} = \mathbfss{C}_T$ in Equation~\eqref{eq:mixmean}. The parameters $m_k$ control the widths of each component of the mixture, so we could imagine each component specifying the distribution of a particular block of the full covariance matrix. However, the mean of each component $\bar{\mathbfss{C}}_k$ must still be symmetric and positive definite, which places restrictions on how we can choose $\bar{\mathbfss{C}}_k$.

For example, consider the case $N=2$, and divide the covariance matrix up into four blocks, labelled $\mathbfss{C}_{(1,1)}$, $\mathbfss{C}_{(1,2)}$, $\mathbfss{C}_{(2,1)} = \mathbfss{C}_{(1,2)}^\intercal$, and $\mathbfss{C}_{(2,2)}$. Suppose that block $\mathbfss{C}_{(1,1)}$ consists of the covariance of the matter power spectrum on large scales where perturbation theory is accurate, whereas $\mathbfss{C}_{(2,2)}$ is the covariance on small scales. Clearly we'd like to assign more variance (a lower $m$) to $\mathbfss{C}_{(2,2)}$, since our models are less accurate there. The weighted sum of the two $\bar{\mathbfss{C}}_k$ matrices must equal the total model covariance $\mathbfss{C}_T$. We thus have to set $w_1 \bar{\mathbfss{C}}_1$ equal to the theoretical model for $\mathbfss{C}_{(1,1)}$ and $w_2 \bar{\mathbfss{C}}_2$ to the model for $\mathbfss{C}_{(2,2)}$, with $m_1$ dictating the variance of $\mathbfss{C}_{(1,1)}$ and $m_2$ the variance of $\mathbfss{C}_{(2,2)}$. However, both $\bar{\mathbfss{C}}_1$ and $\bar{\mathbfss{C}}_2$ need to be $p \times p$ matrices, which suggests that we need to partition each $\bar{\mathbfss{C}}_k$ into four blocks, with the $(2,2)$ block of $\bar{\mathbfss{C}}_1$ and the $(1,1)$ block of $\bar{\mathbfss{C}}_2$ set close to zero. This however makes the determinant of the full $\bar{\mathbfss{C}}_k$ matrix close to zero. It is easy to show that the weights $w_k'$ and $w_k''$ entering into the data covariance in Equation~\eqref{eq:Cymix} and the parameter variance in Equation~\eqref{eq:Astatsmix} are both proportional to $\lvert \bar{\mathbfss{C}}_k \rvert^{m_k/2}$. We are thus not at liberty to enforce that each mixture component only contributes significantly to a particular block of the full covariance matrix.
%
%The off-diagonal blocks of the $\bar{\mathbfss{C}}_k$ must have a weighted sum giving the total theoretical value of $\mathbfss{C}_{(1,2)}$, with the additional constraint that the $\bar{\mathbfss{C}}_k$ are both symmetric and positive definite. This latter condition implies that the almost-zero diagonal blocks cannot be exactly zero, and that the off-diagonal sub-blocks cannot be too large. To see this, note that a matrix $\mathbfss{C}$ is positive definite if and only if its leading principal sub-matrix $\mathbfss{C}_{(1,1)}$ is positive definite and the matrix $\mathbfss{C}_{(2,2)} - \mathbfss{C}_{(1,2)}^\intercal  \mathbfss{C}_{(1,1)}^{-1} \mathbfss{C}_{(1,2)}$ is positive definite. So if we require $\bar{\mathbfss{C}}_1$ for example to have negligible (but non-zero) contribution from its $(2,2)$ sub-block, this requires that $\mathbfss{C}_{(1,2)}$ must also be negligible. We can ensure this by enforce this by setting $\mathbfss{C}_{(1,2)} = 0$ for both $\bar{\mathbfss{C}}_1$ and $\bar{\mathbfss{C}}_2$, and setting the sub-leading diagonal blocks to be proportional to some small parameter $\epsilon$, taking the limit $\epsilon \rightarrow 0$ after the marginalizations have been performed. These considerations imply that the IW mixture prior requires the theoretical model covariance to \emph{block diagonal}. This is a limitation in the case of multiple redshift-bins or multiple summary statistics since the covariance matrices in these cases are typically not block diagonal, but it should be reasonable for a single summary statistic at a single redshift.

Could we specify independent IW distributions for each block of the full covariance matrix? We could certainly do this, but the prior distribution of $\mathbfss{C}$ would then not be IW, and we could not then perform the marginalization analytically.

While it does not appear that we can use the IW mixture prior to straightforwardly control the variances of covariance matrix sub-blocks, its high degree of flexibility and analytic marginalization properties render it deserving of further study, which we defer to a future work.

\section{Distribution of the test statistic}
\label{app:chi2dist}

In this section we prove that the sampling distribution of $\chi^2$ as defined in Equation~\eqref{eq:chi2} with $\mathbfss{C}_T = \mathbfss{C}_0$ is independent of the unknown quantities $\boldsymbol{\mu}_0$ and $\mathbfss{C}_0$. Firstly, we note that $\chi^2$ may be written as $\tilde{\mathbfit{y}}^\intercal \tilde{\mathbfss{C}}^{-1}_{\mathbfit{y}}\tilde{\mathbfit{y}}$, where a tilde denotes projection onto the $p-1$ dimensional hypersurface orthogonal to $\boldsymbol{\mu}_0$. The projected data covariance may be written as
\begin{equation}
  \tilde{\mathbfss{C}}_{\mathbfit{y}} = (1-\lambda)\hat{\tilde{\mathbfss{C}}} + \lambda \tilde{\mathbfss{C}}_0,
\end{equation}
where
\begin{equation}
  \hat{\tilde{\mathbfss{C}}} \sim W_{p-1}[\tilde{\mathbfss{C}}_0/(n-1), n-1],
\end{equation}
which follows from the properties of Wishart distributions (e.g. \citealt{Gupta}). By the definition of the sample covariance matrix we have then that
\begin{align}
  &(n-1)\hat{\tilde{\mathbfss{C}}} \sim \sum_{\alpha=1}^{n-1} \mathbfit{Z}_\alpha \mathbfit{Z}^\intercal_\alpha, \nonumber \\
  &\mathbfit{Z}_\alpha \sim N_{p-1}(\mathbf{0},\tilde{\mathbfss{C}}_0).
\end{align}
Therefore the data covariance is distributed as
\begin{equation}
  (n-1)\tilde{\mathbfss{C}}_{\mathbfit{y}} \sim (1-\lambda)\sum_{\alpha=1}^{n-1} \mathbfit{Z}_\alpha \mathbfit{Z}^\intercal_\alpha + \lambda(n-1)\tilde{\mathbfss{C}}_0.
\end{equation}
Now, let $\mathbfss{D}$ be a non-singular matrix such that $\mathbfss{D}\tilde{\mathbfss{C}}_0 \mathbfss{D}^\intercal = \mathbfss{I}$, and define $\mathbfit{y}^* = \mathbfss{D}\tilde{\mathbfit{y}}$ and $\mathbfss{C}^*_{\mathbfit{y}} = \mathbfss{D}\tilde{\mathbfss{C}}_{\mathbfit{y}} \mathbfss{D}^\intercal$. Since $\tilde{\mathbfit{y}} \sim N_{p-1}(\mathbf{0},\tilde{\mathbfss{C}}_0)$ this implies that $\mathbfit{y}^* \sim N_{p-1}(\mathbf{0},\mathbfss{I})$. Then we have
\begin{align}
  \chi^2 &=  \tilde{\mathbfit{y}}^\intercal \tilde{\mathbfss{C}}^{-1}_{\mathbfit{y}}\tilde{\mathbfit{y}}, \nonumber \\
  &= \mathbfit{y}^{*\intercal}\mathbfss{C}^{*-1}_{\mathbfit{y}} \mathbfit{y}^*
  \label{eq:chi2samp}
\end{align}
with $\mathbfss{C}^{*}_{\mathbfit{y}}$ distributed as
\begin{equation}
  (n-1)\mathbfss{C}^*_{\mathbfit{y}} \sim (1-\lambda)\sum_{\alpha=1}^{n-1} \mathbfit{Z}^*_\alpha \mathbfit{Z}^{*\intercal}_\alpha + \lambda(n-1)\mathbfss{I},
  \label{eq:Cystar}
\end{equation}
where $\mathbfit{Z}^*_\alpha = \mathbfss{D}\mathbfit{Z}_\alpha$ and $\mathbfit{Z}^*_\alpha \sim N_{p-1}(\mathbf{0},\mathbfss{I})$. The distribution of $\chi^2$ is hence completely determined by the quantities $\mathbfit{y}^*$ and $\mathbfit{Z}^*_\alpha$, which together consist of $n(p-1)$ independent standard normal variates. This completes the proof that $\chi^2$ has a distribution independent of $\boldsymbol{\mu}_0$ and $\mathbfss{C}_0$, if and only if $\mathbfss{C}_T = \mathbfss{C}_0$. Equation~\eqref{eq:chi2samp} also tells us how to draw realizations of $\chi^2$. We simply need to draw $n(p-1)$ independent standard normal variates, form the matrix sum in Equation~\eqref{eq:Cystar}, then form the quadratic form in Equation~\eqref{eq:chi2samp}. The cumulative distribution can be formed numerically from these samples and a look-up table produced, from which we may interpolate to compute the PTE of any measured vale of $\chi^2$.

\section{Approximate formula for the average amplitude variance with the Jeffreys' prior}
\label{app:varAJapprox}

With a Jeffreys' prior, the posterior variance on $A$ in Equation~\eqref{eq:Astats} becomes
\begin{equation}
  \mathrm{var}(A) = \frac{n-1}{n-3}\left(\boldsymbol{\mu}_0^\intercal \hat{\mathbfss{C}}^{-1} \boldsymbol{\mu}_0\right)^{-1}\left(1 + \frac{(p-1)T_J^2}{n-p+1}\right),
\end{equation}
with $T_J^2$ defined in Equation~\eqref{eq:T2J}. The quantity $\left(\boldsymbol{\mu}_0^\intercal \hat{\mathbfss{C}}^{-1} \boldsymbol{\mu}_0\right)^{-1}$ is distributed as $\mathrm{Gamma}\left(\frac{n-p}{2},\frac{1}{2q}\right)$ with $q=\left(\boldsymbol{\mu}_0^\intercal \mathbfss{C}_0^{-1} \boldsymbol{\mu}_0\right)^{-1}/(n-1)$, and hence $\langle  \left(\boldsymbol{\mu}_0^\intercal \hat{\mathbfss{C}}^{-1} \boldsymbol{\mu}_0\right)^{-1} \rangle = \frac{n-p}{n-1}\left(\boldsymbol{\mu}_0^\intercal \mathbfss{C}_0^{-1} \boldsymbol{\mu}_0\right)^{-1}$. Since $T_J^2 \sim F_{p-1,n-p+1}$ we have that $\langle T_J^2 \rangle = \frac{n-p+1}{n-p-1}$. To a very good approximation the two terms in parentheses in Equation~\eqref{eq:varAJapprox} are uncorrelated, and so
\begin{equation}
  \langle \mathrm{var}(A) \rangle \approx \frac{(n-2)(n-p)}{(n-3)(n-p-1)}\left(\boldsymbol{\mu}_0^\intercal \mathbfss{C}_0^{-1} \boldsymbol{\mu}_0\right)^{-1}.
  \label{eq:varAJapprox}
\end{equation}
Note that this differs from the average of $\langle 1/F_{AA} \rangle = \frac{n+2}{n}\left(\boldsymbol{\mu}_0^\intercal \mathbfss{C}_0^{-1} \boldsymbol{\mu}_0\right)^{-1}$ since it accounts for the non-Gaussianity of the posterior.
  
%\section{Parameter biases from an incorrect model covariance matrix}
%\label{app:parbias}
%
%In this section we will investigate whether the maximum likelihood estimate (MLE) derived from the marginal likelihood in Equation~\eqref{eq:marglike} is biased when an incorrect model for the covariance matrix is used. Firstly we will assume that only the mean $\boldsymbol{\mu}$ carries parameter dependence. The MLE is defined as the solution to
%
%\begin{align}
%  \frac{\partial}{\partial \theta_\alpha} &= \frac{n+m}{n+m-p-2}\left[ 1 + \frac{(\mathbfit{y}-\boldsymbol{\mu})^\intercal \mathbfss{C}_{\mathbfit{y}}^{-1}(\mathbfit{y}-\boldsymbol{\mu})}{n+m-p-2}\right]^{-1}\frac{\partial \boldsymbol{\mu}^\intercal}{\partial \theta_\alpha} \mathbfss{C}_{\mathbfit{y}}^{-1} (\mathbfit{y}-\boldsymbol{\mu}) \nonumber \\
%  &= 0.
%  \label{eq:logmarglike}
%\end{align}
%
%This is solved by 

%%%%%%%%%%%%%%%%%%%%%%%%%%%%%%%%%%%%%%%%%%%%%%%%%%

% Don't change these lines
\bsp	% typesetting comment
\label{lastpage}
\end{document}